\documentclass[twocolumn,english,showpacs,aps,pre,secnumarabic]{revtex4}
\usepackage[T1]{fontenc}
\usepackage[latin1]{inputenc}
\usepackage{amsmath}
\usepackage{graphicx}
\usepackage{amssymb}

\makeatletter

\providecommand{\tabularnewline}{\\}

\usepackage{graphicx}
\usepackage{amssymb}

\providecommand{\tabularnewline}{\\}

\usepackage{amsfonts}

\usepackage{babel}

\usepackage{babel}
\makeatother
\begin{document}

\title{Universality and quantum effects in one-component critical fluids}

\author{Yves Garrabos }

\affiliation{Equipe du Supercritique pour l'Environnement, les Matériaux et l'Espace
- Institut de Chimie de la Matière Condensée de Bordeaux - Centre
National de la Recherche Scientifique - Université Bordeaux I - 87,
avenue du Docteur Schweitzer, F 33608 PESSAC Cedex France.}

\email{garrabos@icmcb-bordeaux.cnrs.fr}


\date{16 December 2005}

\begin{abstract}
Non-universal scale transformations of the physical fields are extended
to pure quantum fluids and used to calculate susceptibility, specific
heat and the order parameter along the critical isochore of He$^{3}$
near its liquid-vapor critical point. Within the so-called preasymptotic
domain, where the Wegner expansion restricted to the first term of
confluent corrections to scaling is expected valid, the results show
agreement with the experimental measurements and recent predictions,
either based on the minimal-substraction renormalization and the massive
renormalization schemes within the $\Phi_{d=3}^{4}\left(n=1\right)$-model,
or based on the crossover parametric equation of state for Ising-like
systems. 
\end{abstract}
\maketitle

\section{Introduction}

It is well known that the thermodynamic quantities of real pure fluids
close to their gas-liquid critical point (CP) follow the asymptotic
power-law behavior predicted for the 3D Ising-like universality class
in the \emph{asymptotic} critical domain where $\kappa\ll\Lambda_{0}$
\cite{Anisimov2000}. The distance to the critical point is here measured
by the parameter $\kappa$, related to the inverse correlation length
$\xi^{-1}$. $\xi$ characterizes the spatial extent of the diverging
fluctuations of the local density which is related to the order parameter
density of the gas-liquid transition. $\Lambda_{0}$ is a (nonuniversal)
finite wave-number characterizing a discrete microscopic structure
of a fluid with spacing $\Lambda_{0}^{-1}$. So that the critical
thermophysical behavior of the fluid properties occurs when $\xi\Lambda_{0}\gg1$.
Asymptotically close to the critical point, this microscopic parameter
$\Lambda_{0}$ which characterizes each pure fluid, becomes unimportant
when the thermodynamic properties become singular. It means that all
the pure fluids in their asymptotic critical domain obey to the two-scale
universality associated to hyperscaling. Their properties can then
be described by the same reduced equation of state (e.o.s.) and the
same correlation functions, using only two dimensionless parameters
which are two fluid-dependent parameters, in conformity with the two-scale-factor
universality of the 3D Ising-like universality class.

However, it is also now well established that away from this asymptotic
critical region, the properties of real pure fluids can deviate from
hyperscaling. This deviation can take origin on crossover phenomenon
which reflects a competition between universality and nonuniversality
when $\xi\Lambda_{0}\gtrapprox1$. This crossover problem has been
investigated in considerable details, mainly in the classical-to-critical
crossover framework of field theory \cite{ZinnJustin2002}. The resulting
field theoretical crossover functions describe the crossover behavior
of the $\phi_{d=3}^{4}\left(n=1\right)$ model in the universality
class $\mathcal{O}\left(n=1\right)$ in three dimensions ($n=1$ is
the dimension of the order parameter density for the critical transition,
and $d=3$ is the space dimension of the sytem). A better understanding
of non-universal behavior linked to finite values (although large)
of the correlation length is then accounted for by a restricted summation
of the Wegner expansion \cite{Wegner1972}, which introduces one additional
system-dependent parameter to characterize the preasymptotic domain
\cite{Bagnuls1984a,Bagnuls1985} {[}as discussed in \cite{Bagnuls1985}
the singular power laws expressed at the first-order of the Wegner
expansion are expected to be valid within the preasymptotic domain{]}.
Moreover, the values of the adjustable parameters can then be dependent
on the approximations needed by each particular renormalization scheme.
As a practical result, the microscopic length, the crossover parameter,
as well as the two asymptotic scale factors, enter in a larger set
of \emph{adjustable} parameters, including obviously a extended set
of \emph{measurable} critical point coordinates. Therefore the exact
nature of the two asymptotic scale factors for the fluid physical
fields, still remains an implicit open question. This is still the
object of a debating situation \cite{Kim2003}, due to the fact that
\emph{fluid variables} have no definite critical scaling dimensionality
\textit{at finite distance} to the critical point.

The asymptotic existence of such two scale factors proper to the one-component
fluid subclass, were initially postulated in \cite{Garrabos1982,Garrabos1985}
on a phenomenological basis supporting the asymptotic results of the
massive renormalization scheme \cite{Bagnuls1984a,Bagnuls1985,Bagnuls1987,Bagnuls1984b}
of the field theory framework. It was hypothetized that the complete
information to estimate asymptotic singular fluid behaviors is provided
by the \emph{experimental critical point location}, i.e. by a minimal
parameter set, noted $Q_{c,a_{\bar{p}}}^{min}$ \cite{aparticle notation},
composed of four (generalized) critical coordinates (the subscript
$c$ refers to a property defined at the critical point, while the
subscript $\bar{p}$ refers to a property normalized per particle).
This minimal set defines the critical point location on the equilibrium
phase surface of equation $\Phi(p,v_{\bar{p}},T)=0$, where $p$ is
the pressure, $v_{\bar{p}}=\frac{V}{N}$ is the volume per particle,
and $T$ is the temperature ($N$ is the total number of particles
occupying a total volume $V$). The generalized coordinates are composed
by three usual critical point coordinates and one preferred direction
of the tangent plane to the phase surface. Using xenon as a standard
critical fluid \cite{Garrabos1982,Garrabos1985,Bagnuls1984b}, it
was then proposed to perform adequate scale dilatations of the two
relevant physical variables for each one-component fluid. Applying
such a scale dilatation method, we were able to {}``renormalize''
(i.e. rescale) the physical singular behavior of any one-component
fluid on the corresponding {}``master'' (i.e. unique) singular behavior,
where master (i.e. constant amplitudes) features with respect to the
one-component fluid subclass are conform to universal features with
respect to the complete 3D Ising like universality class.

Specifically, this initial substantiation of master scaling is based
on the explicit choice of the same metric factor for thermodynamics
\textit{and} correlations. That permits an unambiguous definition
of the microscopic length $\Lambda_{0}^{-1}$ proportional to a critical
length scale factor $\alpha_{c}=\left(\frac{k_{B}T_{c}}{p_{c}}\right)^{\frac{1}{d}}$
made from an appropriate combination of the critical temperature $T_{c}$
and pressure $p_{c}$ coordinates {[}$k_{B}$ is the Boltzmann constant
and $d=3${]}. From well-known shorted-range of the Lennard-Jones
(LJ) like molecular interactions in one-component fluids \cite{Hirschfelder1954},
characterized by the equilibrium position $r_{e}^{LJ}$ between two
interacting particles, we have $\Lambda_{0}^{-1}=\alpha_{c}\cong2r_{e}^{LJ}$
\cite{Garrabos1982}, ignoring then the possible contribution of quantum
effects on $\Lambda_{0}^{-1}$. In this paper, using the recent experimental
measurements near the critical point of $^{3}He$ \cite{Hahn2001,Zhong2003},
we extend this scale dilatation method (SDM) to the quantum fluid
case. This extension is based on a phenomenological modification of
the non-quantum renormalized critical behavior, which is only valid
at the critical temperature. Since experimental values of the minimum
critical set already contain their actual contribution of quantum
effects, we expect that remaining part of quantum effects only affect
the microscopic length $\left(\Lambda_{0}\right)^{-1}$, in such a
relative way that $\left(\Lambda_{0}\Lambda_{qe}^{*}\right)^{-1}=\alpha_{c}$
{[}see below our Eq. (\ref{wavelength vs alphac}){]}. The adjustable
dimensionless parameter $\Lambda_{qe}^{*}$ is here introduced in
order to maintain the master features observed for the one-component
fluid subclass. Therefore, in addition to the minimal set of four
critical parameters, the renormalized variables need to use only one
supplementary well-defined dimensionless parameter $\Lambda_{qe}^{*}$,
whose value is, either fluid-independent ($\Lambda_{qe}^{*}=1$) in
the absence of quantum effects, or quantum-fluid-dependent ($\Lambda_{qe}^{*}>1$)
in the presence of quantum effects, without violating the asymptotic
universal features of the 3D Ising-like universality class.

The paper is organized as follows. In section 2 we recall the basic
elements of the scale dilatation method and we introduce its extension
to account for quantum effects on the microscopic length scale. In
section 3 we consider the fitting results \cite{Zhong2003} obtained
by Zhong et al for $^{3}He$ to discuss our estimated value of the
adjustable parameter $\Lambda_{qe}^{*}\left(^{3}He\right)=1.11009$.
Before to conclude, the section 4 gives a brief comparison with three
crossover modelling of $^{3}He$ critical properties.

\section{Scale dilatation of the fluid physical variables}

\subsection{The minimal set $Q_{c,a_{\bar{p}}}^{min}$ of four scale factors}

As recalled in our introduction, the basic idea \cite{Garrabos1982,Garrabos1985}
of the scale dilatation method relies on a simple thermodynamic assertion
concerning the thermodynamic information provided by the critical
point location on the fluid phase surface of equation $\Phi(p,v_{\bar{p}},T)=0$
\cite{aparticle notation}. The minimum of information needed to predict
singular thermodynamic behavior of a pure fluid is given by: 

(1) the three critical coordinates $T_{c}$, $p_{c}$, and $v_{\overline{p},c}$
of the liquid-vapor critical point;

(2) the two preferred directions which define the position of the
tangent plane to the phase surface at the critical point (both needed
in order to characterize the linearized asymptotic approach towards
the critical point along two well-defined thermodynamic paths). One
direction is common to all pure fluids (since $\left(\frac{\partial p}{\partial v_{\bar{p}}}\right)_{T=T_{c}}=\left(\frac{\partial T}{\partial v_{\bar{p}}}\right)_{p=p_{c}}=0$),
and only the second direction,\begin{equation}
\gamma_{c}^{'}=\left(\frac{\partial p}{\partial T}\right)_{v_{\bar{p}}=v_{\bar{p},c}}=\left(\frac{dp_{sat}}{dT}\right)_{T=T_{c}}\label{gammac slope}\end{equation}
 is characteristic of each pure fluid. $p_{sat}$ is the saturation
pressure in the non homegeneous domain. We note \begin{equation}
Q_{c,a_{\bar{p}}}^{min}=\left\{ T_{c},p_{c},v_{\overline{p},c},\gamma_{c}^{'}\right\} \label{minimum critical set}\end{equation}
 this minimal set made of four critical parameters.

From thermodynamic principles, this topological information concerns
all the incipient equilibrium states very close to the unstable single
critical point.

From these four coordinates we can calculate unequivocally the following
four fluid characteristic parameters,\begin{equation}
\left(\beta_{c}\right)^{-1}=k_{B}T_{c}\sim\left[energy\right]\label{energy unit}\end{equation}

\begin{equation}
\alpha_{c}=\left(\frac{k_{B}T_{c}}{p_{c}}\right)^{\frac{1}{d}}\sim\left[length\right]\label{length unit}\end{equation}
\begin{equation}
Z_{c}=\frac{p_{c}m_{\bar{p}}}{\rho_{c}k_{B}T_{c}}\label{isothermal compression factor}\end{equation}

\begin{equation}
Y_{c}=\left(\gamma_{c}^{'}\frac{T_{c}}{P_{c}}\right)-1\label{isochoric factor}\end{equation}
 where $\rho_{c}=\left(\frac{N}{V}\right)_{c}m_{\bar{p}}$ is the
critical density of the system made of particles of known individual
mass $m_{\bar{p}}$. $\left(\beta_{c}\right)^{-1}$of Eq. (\ref{energy unit})
fixes the energy unit at the macroscopic scale. $\alpha_{c}$ of Eq.
(\ref{length unit}) fixes the length unit at the macroscopic scale.
The two Eqs. (\ref{energy unit},\ref{length unit}) are sufficient
to make dimensionless all the thermodynamic and correlation functions
of pure fluids \cite{Garrabos1982,Garrabos1985}. $Z_{c}$ of Eq.
(\ref{isothermal compression factor}) is the critical compression
factor. We then introduce the useful compression factor $Z=\frac{-J\left(T,V,\mu_{\bar{p}}\right)}{k_{B}T}$
as the dimensionless opposite form of the total Grand potential $J\left(T,V,\mu_{\bar{p}}\right)=-p\left(T,\mu_{\bar{p}}\right)\times V$,
expressed in terms of its three natural intensive variables $T$,
$V$, and $\mu_{\bar{p}}$ {[}For the total system, $\mu_{\bar{p}}$
is the chemical potential per particle, i.e. the intensive variable
conjugated to $N$, independent to $p$ and $T$, respectively, which
are the two other independent intensive variables, conjugated to $V$
and $S$, respectively{]}. From the experimental phase surface of
equation $\Phi\left(p,v_{\bar{p}},T\right)=0$ it is easy to construct
another practical phase surface of equation $\Phi\left(Z,\rho,T\right)=0$.
In such a representation of the fluid equilibrium states, the characteristics
numbers $Z_{c}$ of Eq. (\ref{isothermal compression factor}) and
$Y_{c}Z_{c}$ made of the product between Eqs. (\ref{isothermal compression factor})
and (\ref{isochoric factor}), read as follows\begin{equation}
Z_{c}=-\left[\left(\frac{\partial Z}{\partial\rho}\right)_{\varpi}\right]_{CP}=-\left[\left(\frac{\partial Z}{\partial\rho}\right)_{LVE}\right]_{CP}\label{isothermal preferred direction}\end{equation}
\begin{equation}
Y_{c}Z_{c}=\left[\left(\frac{\partial Z}{\partial T}\right)_{\rho}\right]_{CP}\label{isochoric preferred direction}\end{equation}
where all the derivatives refer to their values for the critical point
coordinates, while $\varpi$ means any isocline at constant (critical
) value of one intensive variable $\varpi$ among $T$, $p$, or $\mu_{\bar{p}}$,
and $LVE$ means the liquid vapor equilibrium line. Therefore, the
two caracteristic numbers $Z_{c}$ and $Y_{c}Z_{c}$ are the two {}``preferred''
critical directions \cite{preferred} at the critical point of the
phase surface, for the critical isotherm path and the critical isochore
path, respectively. 

From basic modeling of a binary effective interaction characterized
by a minimum energy well depth $\varepsilon_{m}^{LJ}$ at the pair
equilibrium position $r_{m}^{LJ}$ between two particles, we obtain,
$\left(\beta_{c}\right)^{-1}\cong\varepsilon_{m}^{LJ}$ and $\alpha_{c}\cong2$
$r_{m}^{LJ}$, where $\varepsilon_{m}^{LJ}$ and $r_{m}^{LJ}$ are
the respective natural units for energy and length, at the microscopic
scale. Here the subscript $LJ$ stands for a \textit{short-ranged}
Lennard-Jones-like \textit{potential} \cite{Hirschfelder1954}. It
follows that $\alpha_{c}$ measures the mean extension range of the
attractive dispersion forces and\begin{equation}
v_{c,I}=\left(\alpha_{c}\right)^{d}\label{CIC volume}\end{equation}
is the critical volume of \textit{\emph{the microscopic}} \textit{critical
interaction cell}. In such a configuration, the inverse of the critical
compression factor takes clear physical meaning since\begin{equation}
\frac{1}{Z_{c}}=\frac{v_{c,I}}{v_{\overline{p},c}}=n_{c,I}^{\ast}\label{CIC particle number}\end{equation}
 is the \textit{number of fluid particles filling the interaction
cell at criticality,} i.e. for $T=T_{c}$, $n=n_{c}$, and $\Lambda_{0}\xi=\infty$
{[}$v_{\overline{p},c}=(\frac{V}{N})_{PC}$ is the critical volume
per particle, and $n$ ($n_{c}$) is the (critical) number density{]}.
From that result, to formulate dimensionless thermodynamics in terms
of normalization per particle (subscript $\overline{p}$), or in terms
of normalization per critical intercation cell (subscript $I$), appears
easy. As an immediate consequence, from Eqs. (\ref{CIC particle number})
and (\ref{isochoric preferred direction}), $\frac{1}{Z_{c}}$ and
$Y_{c}$ are two characteristics numbers of the critical interaction
cell.

The next step consists to postulate that the two numbers $\left\{ Z_{c};Y_{c}\right\} $,
defined by the two Eqs. (\ref{isothermal compression factor}) and
(\ref{isochoric factor}), are the remaining pair of dimensionless
characteristic parameters at the \emph{scale of the critical interaction
volume,} whatever the selected one-component fluid. In addition, it
is admitted that $Z_{c}$ is the characteristic factor of the scaling
at the critical point and along the critical isotherm, while $Y_{c}$
is the characteristic factor of the scaling along the critical isochore.
Rewriting Eq. (\ref{minimum critical set}) as\begin{equation}
Q_{c,a_{\bar{p}}}^{min}=\left\{ \left(\beta_{c}\right)^{-1},\alpha_{c},Z_{c},Y_{c}\right\} _{CIC}\label{four scale factor set}\end{equation}
we can expect that the complete information is made from four scale
factors which characterize the critical intercation cell (subscript
$CIC$). Then, as initially proposed in \cite{Garrabos1982,Garrabos1985},
the master singular behavior of the correlation functions at exact
criticality and along the critical isochore permits one to link unequivocally
their associated asymptotic amplitudes $\hat{D}$ and $\xi_{0}^{+}$
\cite{Dchapeau}, to $Z_{c}$ and $Y_{c}$, respectively, providing
simultaneously the hyperscaling \cite{Garrabos1985}.

\subsection{Quantum effects on the scale dilatation of physical fluid variables}

The scheme given in \cite{Garrabos1985} also requires that the inverse
microscopic wave number $\Lambda_{0}^{-1}$ is proportional to the
characteristic length scale $\alpha_{c}$. Now, owing to the short
ranged molecular interaction in light pure fluids \cite{Campbell1978},
the influence of quantum mechanical effects changes appreciably the
shape of the Lennard-Jones-like potential, slightly increasing the
range of this interparticle potential \cite{Hirschfelder1954}. This
qualitative evidence was demonstrated by introducing an effective
potential, which is then a temperature-dependent quantity \cite{Thirumalai1983,Pollock1984,Muser2002}.
The quantum effects increase as temperature decreases. However, due
to the formal analogy with the FT renormalization scheme, our rescaling
is basically defined for the critical asymptotic domain, i.e. only
when $T\cong T_{c}$. Moreover, since the use of the actual critical
parameter already includes quantum effects, the remaining additional
quantum effect, for $T\cong T_{c}$, acts only through the relative
modification of the microscopic length at $T_{c}$. In the absence
of theoretical support to do this modification, we propose to normalize
its contribution with respect to the microscopic inverse wave number
defined for non-quantum fluids.

This contribution is expected low and then limited to a small additive
departure from unity. This additive value, noted $\lambda_{c}$, can
then include the two main phenomenological characteristics of quantum
particles :

i) their low mass and size, accounted for using proportionality to
the ratio $\frac{\Lambda_{T,c}}{\alpha_{c}}$ between the critical
thermal wavelength,\[
\Lambda_{T,c}=\frac{h_{P}}{\left(2\pi m_{\bar{p}}k_{B}T_{c}\right)^{\frac{1}{2}}}\]
 (where $h_{P}$ is the Planck constant), and our microscopic critical
range $\alpha_{c}$ of the interaction;

ii) their statistics (like bosons, fermions, etc.), accounted for
by introducing a supplementary free parameter, noted $\lambda_{q,f}$.

So that, we characterize the quantum corrections by the following
non-dimensional factor\begin{equation}
\Lambda_{qe}^{*}=1+\lambda_{c}\label{lambdaqestar}\end{equation}
 with\begin{equation}
\lambda_{c}=\lambda_{q,f}\frac{\Lambda_{T,c}}{\alpha_{c}}\label{lambdaqf vs lambdac}\end{equation}
 $\lambda_{c}\geq0$ is then the measure of the relative modification
of the shape and range of molecular interaction due to the quantum
effects.

Since the quantum effects increase slightly the range of the molecular
interaction, we postulate that the \textit{corrected} microscopic
wave number now reads\begin{equation}
\Lambda_{0}\Lambda_{qe}^{*}=\frac{1}{\alpha_{c}}\label{wavelength vs alphac}\end{equation}
 (in a non-quantum fluid, our previous relation was $\Lambda_{0}=\frac{1}{\alpha_{c}}$,
implicitely). We expect that the rescaled \textit{quantum}-fluid correlation
length $\ell_{qf}^{\ast}$ presents the master divergence previously
defined for all the non-quantum one-component fluids \cite{Garrabos1985,Garrabos2002}.
Then,\begin{equation}
\ell_{qf}^{\ast}=\Lambda_{0}\xi=\left(\Lambda_{qe}^{*}\right)^{-1}\xi^{\ast}\label{lqfstar}\end{equation}
 with\begin{equation}
\xi^{\ast}=\frac{\xi}{\alpha_{c}}\label{ksistar}\end{equation}
 Similarly, any rescaled singular thermodynamic property of the quantum-fluid
can be formulated from dimensional analysis, in order to account for
its proper $\Lambda_{qe}^{*}$ contribution within $v_{c,I}$, which
maintains valid the previous master hypotheses made for the non-quantum
fluid subclass.

Therefore, thanks to the formal analogy between the scale dilatation
method \cite{Garrabos1985} and the basic hypotheses of the renormalization
group approach \cite{Wilson1971,Wilson1974}, all the above quantum
corrections are intrinsically accounted for according our renormalization
scheme, provided that the transformations (dilatations) of the two
relevant physical fields are made throughout the following analytical
relations \begin{equation}
\mathcal{T}_{qf}^{*}\equiv\mathcal{T}^{*}=Y_{c}\Delta\tau^{*}\label{Temperature scale}\end{equation}
\begin{equation}
\mathcal{H}_{qf}^{*}=\left(\Lambda_{qe}^{*}\right)^{2}\mathcal{H}^{*}=\left(\Lambda_{qe}^{*}\right)^{2}\left(Z_{c}\right)^{-\frac{d}{2}}\Delta h^{*}\label{External field scale}\end{equation}
 Consequently, the dilatation of the physical order parameter density
reads as follow \begin{equation}
\mathcal{M}_{qf}^{\ast}=\Lambda_{qe}^{\ast}\mathcal{M}^{\ast}=\Lambda_{qe}^{\ast}\left(Z_{c}\right)^{\frac{d}{2}}\Delta m^{\ast}\label{Order parameter scale}\end{equation}
 In Eqs. (\ref{Temperature scale}) to (\ref{Order parameter scale}),{\small \begin{equation}
\Delta\tau^{\ast}=k_{B}\beta_{c}\left(T-T_{c}\right)\label{reduced temperature distance}\end{equation}
\begin{equation}
\Delta h^{\ast}=\beta_{c}\left(\mu_{\overline{p}}-\mu_{\overline{p},c}\right)\label{reduced ordering field}\end{equation}
}while,\begin{equation}
\Delta m^{\ast}=\left(n-n_{c}\right)\left(\alpha_{c}\right)^{d}\label{reduced OP density}\end{equation}
 $\mu_{\overline{p},c}$ is the critical chemical potential per particle.
Obviously, in Eqs. (\ref{Temperature scale}) to (\ref{Order parameter scale}),
$\mathcal{T}^{*}$, $\mathcal{H}^{*}$ and $\mathcal{M}^{*}$ are
the renormalized variables already defined for non-quantum fluids
\cite{Garrabos1985,Garrabos2002}.

\subsection{Master and physical singular behavior}

Because such transformations of the physical fields in the FT framework
have a range of validity including (at least) the first correction-to-scaling
\cite{Bagnuls2002}, our rescaled thermodynamic and correlation functions
should conform to the two-term (leading and first-confluent) asymptotic
description of singularities within the preasymptotic domain. For
example when $\mathcal{T}^{\ast}$ goes to zero along the critical
isochore, the critical behavior of any rescaled singular property
$\mathcal{P}_{qf}^{\ast}$ reads \begin{equation}
\mathcal{P}_{qf}^{\ast}=\mathcal{Z}_{\mathcal{P}}^{\pm}\left|\mathcal{T}^{\ast}\right|^{-x}\left[1+\mathcal{Z}_{\mathcal{P}}^{1,\pm}\left|\mathcal{T}^{\ast}\right|^{\Delta}+\mathcal{O}\left(\left|\mathcal{T}^{\ast}\right|^{2\Delta}\right)\right]\label{P Master two term power law}\end{equation}
 where $x$ and $\Delta$ are the associated universal critical exponents
\cite{Guida1998}. The subscript $+$ is for the homogenous domain
$\mathcal{T}^{\ast}>0$ (i.e. $T>T_{c}$), and the subscript $-$
is for the non-homogeneous domain $\mathcal{T}^{\ast}<0$ (i.e. $T<T_{c}$).
The leading amplitudes $\mathcal{Z}_{\mathcal{P}}^{\pm}$, and the
first confluent amplitudes $\mathcal{Z}_{\mathcal{P}}^{1,\pm}$, are
\textit{master} (constant) numbers for all pure fluids. Their respective
values are obtained using xenon as a standard critical fluid \cite{Garrabos2002},
and accounting as closely as possible for up-to-date estimates \cite{Guida1998,Bagnuls2002}
of universal asymptotic critical quantities (exponents and amplitudes
combinations). 

When the generalized critical parameters of a pure fluid are known,
there is an immediate practical interest to reverse the use of the
scale dilatation method. In fact, the basic advantage of this method
is its ability to calculate all the amplitudes appearing in the singular
divergences expressed at first-order of the Wegner expansion in $\Delta\tau^{\ast}$.
For $\Delta\tau^{\ast}\rightarrow0^{\pm}$, the critical behavior
of the physical property $P$ of the selected pure fluid is represented
by the two-term equation

\begin{equation}
P=P_{0}^{\pm}\left|\Delta\tau^{\ast}\right|^{-x}\left[1+P_{1}^{\pm}\left|\Delta\tau^{\ast}\right|^{\Delta}+\mathcal{O}\left(\left|\Delta\tau^{\ast}\right|^{2\Delta}\right)\right]\label{P fluid two term power law}\end{equation}
 where $P_{0}^{\pm}$ and $P_{1}^{\pm}$ are the leading and the first
confluent amplitudes. All the values of $P_{0}^{\pm}$ and $P_{1}^{\pm}$
can then be estimated when the basic set of critical parameters is
known for a selected pure fluid, using each unequivocal relation linking
the physical quantity to its renormalized one (see Section 3 below,
and the Table I, columns 6 and 7).

\begin{center}%
\begin{table*}
\begin{center}\begin{tabular}{|c|c|c|c|c|c|c|}
\hline 
$\mathcal{P}_{qf}^{\ast}$&
 $x$&
 $\mathcal{Z}_{\mathcal{P}}^{\pm}$&
 $\mathcal{Z}_{\mathcal{P}}^{1,\pm}$&
$P$&
 $P_{0}^{\pm}$&
 $P_{1}^{\pm}$\tabularnewline
\hline
$\ell_{qf}^{\ast,+}$&
 $\begin{array}{c}
\nu=0.6304\pm0.0013\end{array}$&
 $\begin{array}{c}
\mathcal{Z}_{\ell}^{+}=0.570365\end{array}$&
 $\begin{array}{c}
\mathcal{Z}_{\ell}^{1,+}=0.37685\end{array}$&
$\xi$&
 $\begin{array}{c}
\xi_{0}^{+}=\alpha_{c}\Lambda_{qe}\left(Y_{c}\right)^{-\nu}\mathcal{Z}_{\ell}^{+}\end{array}$&
 $\begin{array}{c}
a_{\ell}^{+}=\mathcal{Z}_{\ell}^{1,+}\left(Y_{c}\right)^{\Delta}\end{array}$\tabularnewline
$\chi_{qf}^{\ast,+}$&
 $\begin{array}{c}
\gamma=1.2397\pm0.0013\end{array}$&
 $\begin{array}{c}
\mathcal{Z}_{\chi}=0.119\end{array}$&
 $\begin{array}{c}
\mathcal{Z}_{\chi}^{1,+}=0.555\end{array}$&
$\chi^{*}$&
 $\begin{array}{c}
\Gamma^{+}=\left(\Lambda_{qe}^{*}\right)^{d-2}\left(Z_{c}\right)^{-1}\left(Y_{c}\right)^{-\gamma}\mathcal{Z}_{\chi}^{+}\end{array}$&
 $\begin{array}{c}
a_{\chi}^{+}=\mathcal{Z}_{\chi}^{1,+}\left(Y_{c}\right)^{\Delta}\end{array}$\tabularnewline
$\mathcal{C}_{qf}^{\ast,+}$&
 $\begin{array}{c}
\alpha=0.1088\pm0.0039\end{array}$&
 $\begin{array}{c}
\mathcal{Z}_{\mathcal{C}}^{+}=0.105656\end{array}$&
 $\begin{array}{c}
\mathcal{Z}_{\mathcal{C}}^{1,+}=.52310\end{array}$&
$\Delta c_{V}^{*}$&
 $\begin{array}{c}
\frac{A^{+}}{\alpha}=\left(\Lambda_{qe}^{*}\right)^{-d}\left(Y_{c}\right)^{2-\alpha}\mathcal{Z}_{\mathcal{C}}^{+}\end{array}$&
 $\begin{array}{c}
a_{C}^{+}=\mathcal{Z}_{\mathcal{C}}^{1,+}\left(Y_{c}\right)^{\Delta}\end{array}$\tabularnewline
$\mathcal{M}_{qf}^{*}$&
 $\begin{array}{c}
-\beta=-0.3258\pm0.0014\end{array}$&
 $\begin{array}{c}
\mathcal{Z}_{\mathcal{M}}=0.468\end{array}$&
 $\begin{array}{c}
\mathcal{Z}_{\mathcal{M}}^{1}=0.4995\end{array}$&
$\Delta\rho_{LV}^{*}$&
 $\begin{array}{c}
B=\left(\Lambda_{qe}^{*}\right)^{-1}\left(Z_{c}\right)^{-\frac{1}{2}}\left(Y_{c}\right)^{\beta}\mathcal{Z}_{\mathcal{M}}\end{array}$&
 $\begin{array}{c}
a_{M}=\mathcal{Z}_{\mathcal{M}}^{1}\left(Y_{c}\right)^{\Delta}\end{array}$ \tabularnewline
\hline
\end{tabular}\end{center}

\caption{Parameters for the master critical behavior of the correlation length,
the susceptibility, the specific heat and the liquid-gas coexisting
density along the critical isochore of pure fluid. Exponent values
on column 2, amplitude ratios values, and $\Delta=0.502\pm0.004$,
are from \cite{Bagnuls2002,Guida1998}.}
\end{table*}
\end{center}

\section{The helium 3 case}

\subsection{Notations}

The scale dilation method is now applied to the description of the
isothermal susceptibility, specific heat, and coexisting density measurements
\cite{Zhong2003} along the critical isochore of $^{3}He$. We complete
these measurements by the estimation of the correlation length inferred
from the two-scale-factor universality. Let us introduce the corresponding
notations for: 

i) the master singular behaviors\begin{equation}
\mathcal{\ell}_{qf}^{\ast}=\mathcal{Z}_{\mathcal{\ell}}^{\pm}\left(\mathcal{T}^{\ast}\right)^{-\nu}\left[1+\mathcal{Z}_{\mathcal{\ell}}^{1,\pm}\left(\mathcal{T}^{\ast}\right)^{\Delta}\right]\label{lqf master two term power law}\end{equation}
\begin{equation}
\mathcal{\chi}_{qf}^{\ast}=\mathcal{Z}_{\mathcal{\chi}}^{\pm}\left(\mathcal{T}^{\ast}\right)^{-\gamma}\left[1+\mathcal{Z}_{\mathcal{\chi}}^{1,\pm}\left(\mathcal{T}^{\ast}\right)^{\Delta}\right]\label{ksiqf master two term power law}\end{equation}
\begin{equation}
\mathcal{\Delta C}_{qf}^{\ast}=\mathcal{Z}_{\mathcal{C}}^{\pm}\left(\mathcal{T}^{\ast}\right)^{-\alpha}\left[1+\mathcal{Z}_{\mathcal{C}}^{1,\pm}\left(\mathcal{T}^{\ast}\right)^{\Delta}\right]+\label{cqf master two term power law}\end{equation}
\begin{equation}
\mathcal{M}_{qf}^{\ast}=\mathcal{Z}_{\mathcal{M}}\left(\mathcal{T}^{\ast}\right)^{\beta}\left[1+\mathcal{Z}_{\mathcal{M}}^{1}\left(\mathcal{T}^{\ast}\right)^{\Delta}\right]\label{mqf master two term power law}\end{equation}
The universal values of the critical exponents $\nu$, $\gamma$,
$\alpha$, and $\beta$, estimated by Guida et al \cite{Guida1998}
are given in column 2 of Table 1. $\Delta=0.502\left(\pm0.004\right)$
\cite{Guida1998} is the lowest value of the confluent exponent. The
master (i.e. constant) values of the leading ($\mathcal{Z}_{\mathcal{\ell}}^{+}$,
$\mathcal{Z}_{\mathcal{\chi}}^{+}$, $\mathcal{Z}_{\mathcal{C}}^{+}$,
and $\mathcal{Z}_{\mathcal{M}}$) and confluent amplitudes ($\mathcal{Z}_{\mathcal{\ell}}^{1,+}$,
$\mathcal{Z}_{\mathcal{\chi}}^{1,+}$, $\mathcal{Z}_{\mathcal{C}}^{1,+}$,
and $\mathcal{Z}_{\mathcal{M}}^{1}$) are given in columns 3 and 4
(respectively), of Table 1. The master correlation length $\ell_{qf}^{\ast}$
of Eq. (\ref{P Master two term power law}) provides a direct comparison
from the size of the critical fluctuations to the range of molecular
interaction, in order to control that the basic condition $\ell_{qf}^{\ast}\gg1$
for critical phenomena understanding is valid. That provides also
a criteria to define the master extension of the preasymptotic domain
for the one-component fluid subclass \cite{Garrabos2002}. 

ii) the physical singular behaviors

\begin{equation}
\xi=\xi_{0}^{\pm}\left|\Delta\tau^{\ast}\right|^{-\nu}\left[1+a_{\xi}^{\pm}\left|\Delta\tau^{\ast}\right|^{\Delta}\right]\label{Ksi fluid two term power law}\end{equation}
\begin{equation}
\chi_{\rho}^{*}=\Gamma^{\pm}\left|\Delta\tau^{\ast}\right|^{-\gamma}\left[1+a_{\chi}^{\pm}\left|\Delta\tau^{\ast}\right|^{\Delta}\right]\label{kistar fluid two term power law}\end{equation}
\begin{equation}
\Delta c_{V,\rho}^{*}=\frac{A^{\pm}}{\alpha}\left|\Delta\tau^{\ast}\right|^{-\alpha}\left[1+\alpha a_{\chi}^{\pm}\left|\Delta\tau^{\ast}\right|^{\Delta}\right]+B_{cr}^{*}\label{Deltacvstar fluid two term power law}\end{equation}
\begin{equation}
\Delta\widetilde{\rho}_{LV}=B\left|\Delta\tau^{\ast}\right|^{\beta}\left[1+a_{m}\left|\Delta\tau^{\ast}\right|^{\Delta}\right]\label{deltarhostar fluid two term power law}\end{equation}
$\xi$ of Eq. (\ref{Ksi fluid two term power law}) is the correlation
length, i.e. the actual size of the critical fluctuations of the order
parameter density. The Eqs. (\ref{kistar fluid two term power law})
to (\ref{deltarhostar fluid two term power law}) are written with
usefull variables of fluid related critical phenomena \cite{Sengers1986},
which needs a complementary analysis made in the next subsection,
to precise the normalization of the thermodynamics and the role of
the energy and length scale units given by Eqs. (\ref{energy unit})
and (\ref{length unit}), respectively.

\subsection{Thermodynamic properties}

\subsubsection{The isothermal susceptibility}

Considering a mass unit of the fluid as in the standard thermodynamic
presentation of \emph{specific} properties, the susceptibility $\chi_{\rho}=\left(\frac{\partial\rho}{\partial\mu_{\rho}}\right)_{T}=\rho\left(\frac{\partial\rho}{\partial p}\right)_{T}\sim\left[kg^{2}\, J^{-1}\, m^{-3}\right]$
is expressed in units of $\frac{\rho_{c}^{2}}{p_{c}}$, while the
subscript $\rho$ recalls for the thermodynamic normalization per
mass unit. Therefore, in Eq. (\ref{kistar fluid two term power law}),
$\chi_{\rho}^{*}=\chi\frac{p_{c}}{\rho_{c}^{2}}=\kappa_{T}\left(\frac{\rho}{\rho_{c}}\right)^{2}p_{c}=\left(\widetilde{\rho}\right)^{2}\kappa_{T}^{*}$,
with $\kappa_{T}=\frac{1}{\rho}\left(\frac{\partial\rho}{\partial p}\right)_{T}$
and $\kappa_{T}^{*}=p_{c}\kappa_{T}$. $\mu_{\rho}=\frac{\mu_{\overline{p}}}{m_{\overline{p}}}$
is the chemical potential per mass unit, dual from the (mass) density
$\rho$. $\kappa_{T}$ is the isothermal compressibility. $\widetilde{\rho}=\frac{\rho}{\rho_{c}}$
is the practical dimensionless form of the density, which differs
by a factor $Z_{c}$ from the dimensionless form $\rho^{*}=\rho\frac{\left(\alpha_{c}\right)^{d}}{m_{\overline{p}}}$
obtained with our length unit $\alpha_{c}$ \cite{Garrabos1982,Garrabos1985}.
We note that the above susceptibility $\chi_{\rho}$ also differs
from the susceptibility $\chi_{\bar{p}}=\left(\frac{\partial n}{\partial\mu_{\bar{p}}}\right)_{T}=n\left(\frac{\partial n}{\partial p}\right)_{T}\sim\left[J\, m^{3}\right]^{-1}$
where the subscript $\bar{p}$ recalls for the thermodynamic normalization
\emph{per particle}. Expressing $\chi_{\bar{p}}$ in coherent {[}i.e.
using Eqs. (\ref{energy unit}) and (\ref{length unit}){]} units
of $\frac{\beta_{c}}{\left(\alpha_{c}\right)^{d}}$, we obtain $\chi_{\bar{p}}^{*}=\left(\frac{1}{Z_{c}}\right)^{2}\chi_{\rho}^{*}=\left(n_{c}^{*}\right)^{2}\kappa_{T}^{*}$
{[}using $n^{*}=n\left(\alpha_{c}\right)^{d}${]}. However, pressure
($\sim\left[\frac{energy}{volume}\right]$) appears appropriately
expressed in units of $p_{c}=\frac{\left(\beta_{c}\right)^{-1}}{\left(\alpha_{c}\right)^{d}}$,
within the both (practical and coherent) dimensionless formulations.

\subsubsection{The heat capacity at constant volume}

The total heat capacity at constant volume $C_{V}\sim\left[J\, K^{-1}\right]$
of the fluid mass $M$ is divided by the total fluid volume $V$ to
have a unit of $\rho c_{V,\rho}$, where $c_{V,\rho}=\frac{C_{V}}{M}\sim\left[J\, kg^{-1}\, K^{-1}\right]$
is the specific heat at constant volume. The dimensionless specific
heat $c_{V,\rho}^{*}$ is then obtained expressing the total heat
capacity in units of $\frac{p_{c}V}{T_{c}}$, so that $c_{V,\rho}^{*}=\rho c_{V}\frac{T_{c}}{p_{c}}$.
Therefore, in Eq. (\ref{Deltacvstar fluid two term power law}), the
singular specific heat $\Delta c_{V,\rho}^{*}\left(\Delta\tau^{*}\right)$
is such that the total specific heat $c_{V,\rho}^{*}\left(T^{*}\right)$
as a function of $T^{*}=\frac{T}{T_{c}}$ reads as follows\begin{equation}
c_{V,\rho}^{*}\left(T^{*}\right)=\Delta c_{V,\rho}^{*}\left(\Delta\tau^{*}\right)+C_{B,\rho}^{*}\left(T^{*}\right)\label{total cvstar}\end{equation}
In Eq. (\ref{Deltacvstar fluid two term power law}), $B_{cr}^{*}$
is a critical constant while, in Eq. (\ref{total cvstar}), $C_{B,\rho}^{*}\left(T^{*}\right)$
is the regular background reflecting the analytical part of the free
energy. In our coherent formulation of the particle properties, the
heat capacity per particle $c_{V,\bar{p}}=\frac{C_{V}}{N}\sim\left[J\, K^{-1}\right]$
have the (universal) $k_{B}$ dimension. As a matter of fact, the
heat capacity per particle is the unique measurable thermodynamic
property which can be made dimensionless only using the Boltzmann
factor $k_{B}$, i.e. \emph{without reference} to $\alpha_{c}$ and
$\left(\beta_{c}\right)^{-1}$. Therefore, when the singular heat
capacity at constant volume, normalized per particle, obeys the asymptotic
power law\begin{equation}
\Delta c_{V,\bar{p}}=\frac{A_{0,\bar{p}}^{\pm}}{\alpha}\left|\Delta\tau^{*}\right|^{-\alpha}\left[1+\mathcal{O}\left\{ \left|\Delta\tau^{*}\right|^{\Delta}\right\} \right]\label{Wegner eq cvparticle}\end{equation}
 along the critical isochore, \emph{one ($+$ or $-$) among the two
dimensionless amplitudes} $\frac{A_{0,\bar{p}}^{\pm}}{k_{B}}$ \emph{is
mandatorily a characteristic fluid-particle-dependent number} (the
two amplitudes being related by the universal ratio $\frac{A_{0,\bar{p}}^{+}}{A_{0,\bar{p}}^{-}}\approxeq0.537$
for $d=3$ \cite{Guida1998}). However, hyperscaling features impose
that the same length scale is used in thermodynamic and correlation
functions. For example, in the case of an {}``uncompressible'' 3D
Ising-system of the lattice spacing $a_{Ising}$ , the singular part
of the heat capacity normalized by $k_{B}$ can be expressed in unit
of $\left(a_{Ising}\right)^{d}$ \cite{Privman 1991} {[}the extensive
nature of the total number of particle is then implicitely accounted
for in a crystallized solid system since the total volume is proportional
to the cell lattice volume containing a fixed number of particles{]}.
Similarly, in the case of the compressible one-component fluid, that
needs to express normalized heat capacity per particle in unit of
$\left(\alpha_{c}\right)^{d}$ (ignoring in this simple dimensional
analysis the quantum effects on the microscopic wavelength). The number
of particles within the critical interaction cell being $\frac{1}{Z_{c}}$,
we thus define the singular part of the heat capacity for the volume
of the critical interaction cell as follows\begin{equation}
\Delta c_{V,I}^{*}=\frac{1}{Z_{c}}\Delta c_{V,\bar{p}}^{*}\label{dimensionless mcvs}\end{equation}
 whith $\Delta c_{V,\bar{p}}^{*}=\frac{\Delta c_{V,\bar{p}}}{k_{B}}$.
Accordingly, $\frac{1}{Z_{c}}$ takes equivalent microscopic nature
of the coordination number in the lattice description of the three
dimensional Ising systems, while $\alpha_{c}$ takes equivalent microscopic
nature of the lattice spacing $a_{Ising}$. Now, for comparison with
the notations used in fluid-related critical phenomena where all the
thermodynamic potentials are divided by the total fluid volume, we
also introduce the heat capacity at constant volume, \emph{for a fluid
in a container of unit volume,} $\Delta c_{V=1}=\frac{\Delta c_{V,\bar{p}}}{v_{\bar{p},c}}$
(labelled here with the subscript $V=1$). Expressed in our above
unit length scale {[}Eq. (\ref{length unit}){]}, the associated dimensionless
form reads\begin{equation}
\Delta c_{V=1}^{*}=\frac{\Delta c_{V,\bar{p}}}{k_{B}}\times\frac{1}{v_{\bar{p},c}\left(\alpha_{c}\right)^{d}}=\frac{\Delta c_{V=1}}{k_{B}}\times\frac{1}{\left(\alpha_{c}\right)^{d}}\label{deltacvstarVunit}\end{equation}
 Obviously, $\Delta c_{V=1}^{*}$ is identical to

i) the previous dimensionless form $\Delta c_{V,\rho}^{*}=\frac{\Delta C_{V}}{V}\times\frac{T_{c}}{p_{c}}$
of the total singular heat capacity $\Delta C_{V}=N$$\Delta c_{V,\bar{p}}$
of the constant total fluid volume $V$, filled with $N$ (fixed)
particles and,

ii) the our dimensionless form $\Delta c_{V,I}^{*}=\Delta c_{V,\bar{p}}^{*}\times\frac{1}{Z_{c}}$
of the singular heat capacity of the microscopic interacting volume
$v_{c,I}$ {[}Eq. (\ref{CIC volume}){]}, filled with $\frac{1}{Z_{c}}$
(fixed) particles. In this latter situation, we have an explicit comprehension
of the extensive nature of the two independent variables $V$ and
$N$ for compressible fluids. Specially, we note here the importance
of the thermodynamic normalization for better understanding of the
scaling nature of the critical amplitudes, such as in Eq. (\ref{Deltacvstar fluid two term power law})
for example. Considering the hyperscaling law $2-\alpha=d\nu$, throughout
the universal quantity made by the product\[
\left(\Delta\tau^{*}\right)^{2}\times\frac{\Delta c_{V,\rho}^{*}}{\left(\alpha_{c}\right)^{d}}\times\xi^{d}=universal\, quantity\]
and rewriting this product such as\[
\left(\Delta\tau^{*}\right)^{2}\times\frac{1}{Z_{c}}\times\left(\Delta c_{V,\bar{p}}^{*}\right)\times\left(\frac{\xi}{\alpha_{c}}\right)^{d}=\left(R_{\xi}^{\pm}\right)^{d}\]
 {[}with $R_{\xi}^{+}\approxeq0.2696$ and $R_{\xi}^{-}\approxeq0.169$,
for $d=3$ \cite{Guida1998}{]}, we can easily demonstrate that the
universal amplitude combination \begin{equation}
\begin{array}{l}
\left(R_{\xi}^{\pm}\right)^{d}=\frac{1}{Z_{c}}\left(\frac{A_{0,\bar{p}}^{\pm}}{k_{B}}\right)\left(\frac{\xi_{0}^{\pm}}{\alpha_{c}}\right)^{d}\end{array}\label{universal combination Rksi}\end{equation}
 contains the two independent extensive features (volume and number
of particles) of the fluid system at the scale of the critical interaction
cell. In such a situation, the dimensioned leading amplitudes $A_{0,\bar{p}}^{\pm}$
(associated to a particle property), and $\xi_{0}^{\pm}$ (associated
to the microscopic wavelength), have a well-understood physical meaning
with respect to the universal features of the universality class.
As an essential consequence, the universal feature of any singular
free energy must then be expressed in terms of the unique remaining
energy scale $\left(\beta_{c}\right)^{-1}$. We will return below
(see §.3.3) on this important remark to account for quantum effects
in the master singular behavior of the one-component fluid subclass.

\subsubsection{The (dual) densities and chemical potentials}

We finally consider the non homogeneous domain below $T_{c}$, where
the practical dimensionless form of the symmetrized order parameter
density {[}see eq. (\ref{deltarhostar fluid two term power law}){]}
is defined by\begin{equation}
\Delta\widetilde{\rho}_{LV}\left(\left|\Delta\tau^{*}\right|\right)=\frac{\Delta\rho_{LV}\left(\left|\Delta\tau^{*}\right|\right)}{2\rho_{c}}=\frac{\rho_{L}-\rho_{V}}{2\rho_{c}}\label{deltarhotildeLV}\end{equation}
 $\rho_{L}$ ($\rho_{V}$) is the liquid (vapor) density of one coexisting
phase. Such a dimensionless form occurs from the usefull variable
$\widetilde{\rho}=\frac{\rho}{\rho_{c}}$ (see above), leading to
consider the quantity
\begin{equation}
	\Delta\widetilde{\rho}=\frac{\rho-\rho_{c}}{\rho_{c}}=\widetilde{\rho}-1
	\label{practical OP density}
\end{equation}
 as a practical order parameter density, and the quantity\begin{equation}
\Delta\widetilde{\mu}_{\rho}=\frac{\mu_{\rho}-\mu_{\rho,c}}{\mu_{\rho,c}}=\widetilde{\mu}_{\rho}-1\label{practical ordering field}\end{equation}
as a practical ordering field. $\mu_{\rho,c}=\frac{\mu_{\bar{p}}}{m_{\bar{p}}}$
is the specific chemical potential at the critical point. $\widetilde{\mu}_{\rho}=\frac{\mu_{\rho}}{\mu_{\rho,c}}$
is the practical dimensionless form of the chemical potential, which
differs by a factor $\left(\beta_{c}\mu_{\bar{p},c}\right)^{-1}$
from our dimensionless form $\mu_{\bar{p}}^{*}=\beta_{c}\mu_{\bar{p}}^{*}$
obtained with our energy unit $\left(\beta_{c}\right)^{-1}$. From
comparison between the two definitions of the order parameter density
by Eq. (\ref{reduced OP density}) and (\ref{practical OP density}),
we obtain

\begin{equation}
\begin{array}{c}
\Delta\widetilde{\rho}=Z_{c}\Delta m^{\ast}\\
\Delta\widetilde{\rho}_{LV}=Z_{c}\Delta m_{LV}^{\ast}\end{array}\label{deltarhotilde vs deltamstar}\end{equation}
where\begin{equation}
\Delta m_{LV}^{\ast}=\left(n_{L}-n_{V}\right)\left(\alpha_{c}\right)^{d}\label{DeltamLVstar}\end{equation}
 The main conclusive remark to note using these dual variables $\Delta\widetilde{\rho}$
and $\Delta\widetilde{\mu}_{\rho}$, of respective Eqs. (\ref{practical OP density})
and (\ref{practical ordering field}), concerns the implicit addition
of a new length scale factor $\widetilde{\alpha}_{c}=\left(\frac{m_{\bar{p}}}{\rho_{c}}\right)^{\frac{1}{d}}$
and a new energy scale factor $\left(\widetilde{\beta}_{c}\right)^{-1}=m_{\bar{p}}\mu_{\rho,c}$.
As a consequence, the nonuniversal nature of each leading amplitude
is a complex combination of the interrelated dimensionned scale factors
and of the two scale factors associated to universal scaling in fluids.

\subsection{Two-scale-factor universality and quantum effects}

In addition to Eqs. (\ref{lqf master two term power law}) to (\ref{deltarhostar fluid two term power law}),
we now introduce:

i) the renormalized singular free energy density $\mathcal{A}_{qf}^{*}\left(\mathcal{T}^{*},\mathcal{M}_{qf}^{*}\right)$
which, along the isocline $\mathcal{M}_{qf}^{*}=0$, asymptotically
behaves as\begin{equation}
\mathcal{A}_{qf}^{*}\left(\mathcal{T}^{*}\right)=\mathcal{Z}_{\mathcal{A}}^{\pm}\left(\mathcal{T}^{*}\right)^{2-\alpha}\left[1+\mathcal{O}\left\{ \left(\mathcal{T}^{*}\right)^{\Delta}\right\} \right]\label{Aqfstar power law}\end{equation}
 with respect to the master thermal field $\mathcal{T}^{*}$ going
to zero. Correspondingly, the thermodynamics definitions of the renormalized
properties of present interest are $\mathcal{H}_{qf}^{*}\left(\mathcal{T}^{*},\mathcal{M}_{qf}^{*}\right)=\left(\frac{\partial\mathcal{A}_{qf}^{*}}{\partial\mathcal{M}_{qf}^{*}}\right)_{\mathcal{T}^{*}}$,
$\mathcal{\chi}_{qf}^{*}\left(\mathcal{T}^{*},\mathcal{M}_{qf}^{*}\right)=\left(\frac{\partial\mathcal{M}_{qf}^{*}}{\partial\mathcal{H}_{qf}^{*}}\right)_{\mathcal{T}^{*}}$,
$\frac{\mathcal{\Delta C}_{qf}^{*}\left(\mathcal{T}^{*},\mathcal{M}_{qf}^{*}\right)}{\mathcal{T}^{*}}=-\left(\frac{\partial^{2}\mathcal{A}_{qf}^{*}}{\partial\mathcal{T}^{*2}}\right)_{\mathcal{M}_{qf}^{*}=0}$,
{[}with $\mathcal{Z}_{\mathcal{A}}^{\pm}=\frac{\mathcal{Z}_{C}^{\pm}}{\alpha\left(1-\alpha\right)\left(2-\alpha\right)}${]};

ii) the singular part $\Delta a_{\rho}\left(T,\rho\right)=\frac{\Delta A_{\rho}}{V}$
of the Helmholtz free energy density, where temperature $T$ and (practical)
density $\rho=\frac{Nm_{\bar{p}}}{V}$ are the two selected variables
to describe a fluid maintained in a container of constant total volume
$V$. In our case where order parameter density is related to the
(natural) number density $n=\frac{N}{V}$, we note $\Delta a\left(T,n\right)=\frac{\Delta A}{V}$
this singular part of the Helmholtz free energy density. Due to the
appropriate dimensionless form of the pressure mentionned above, both
the usefull dimensionless form $\Delta a_{\rho}^{*}\left(\Delta\tau^{*},\Delta\widetilde{\rho}\right)=\frac{\Delta A_{\rho}}{V}\times\frac{1}{p_{c}}$
and the natural dimensionless form $\Delta a^{*}\left(\Delta\tau^{*},\Delta m^{*}\right)=\beta_{c}\Delta A\times\frac{\left(\alpha_{c}\right)^{d}}{V}$
are identical, except the use of two distinct reduced forms $\Delta\widetilde{\rho}$
and $\Delta m^{*}$ of the order parmeter density. Along the critical
isochore $\Delta\widetilde{\rho}=\Delta m^{*}=0$, the singular part
of the free energy behaves as\begin{equation}
\begin{array}{cl}
\Delta a_{\rho}^{*}\left(\Delta\tau^{*}\right) & \equiv\Delta a^{*}\left(\Delta\tau^{*}\right)\\
 & =A'^{\pm}\left|\Delta\tau^{*}\right|^{2-\alpha}\left[1+\mathcal{O}\left\{ \left|\Delta\tau^{*}\right|^{\Delta}\right\} \right]\end{array}\label{free energy power law}\end{equation}
 The basic thermodynamic definitions of the physical properties are:
$\Delta\widetilde{\mu}_{\rho}\left(\Delta\tau^{*},\Delta\widetilde{\rho}\right)=\left(\frac{\partial\Delta a_{\rho}^{*}}{\partial\Delta\widetilde{\rho}}\right)_{^{\Delta\tau^{*}}}$,
or $\Delta\mu_{\bar{p}}^{*}\left(\Delta\tau^{*},\Delta m^{*}\right)=\left(\frac{\partial\Delta a^{*}}{\partial\Delta m^{*}}\right)_{^{\Delta\tau^{*}}}$;
$\mathcal{\chi}_{\rho}^{*}\left(\Delta\tau^{*},\Delta\widetilde{\rho}\right)=\left(\frac{\partial\Delta\widetilde{\rho}}{\partial\Delta\widetilde{\mu}_{\rho}}\right)_{\Delta\tau^{*}}$,
or $\chi_{\bar{p}}^{*}\left(\Delta\tau^{*},\Delta m^{*}\right)=\left(\frac{\partial\Delta a^{*}}{\partial\Delta\mu_{\bar{p}}^{*}}\right)_{\Delta\tau^{*}}$;
and $\frac{\Delta c_{V,\rho}^{*}\left(\Delta\tau^{*},\Delta\widetilde{\rho}\right)}{T^{*}}=\left(\frac{\partial^{2}\Delta\widetilde{a}_{\rho}}{\partial\left(T^{*}\right)^{2}}\right)_{^{\Delta\widetilde{\rho}}}$,
or $\frac{\Delta c_{V}^{*}\left(\Delta\tau^{*},\Delta m^{*}\right)}{T^{*}}=\left(\frac{\partial^{2}\Delta a^{*}}{\partial\left(T^{*}\right)^{2}}\right)_{^{\Delta\widetilde{\rho}}}$;

It is thus easy to obtain the relations reported in the columns $6$
and $7$ of Table I, using the above basic thermodynamic definitions
of the renormalized and physical variables. That also provides a comprehensive
understanding of the quantum effect correction to master singular
behavior.

As a matter of fact, following the argument first proposed by Widom
\cite{Widom1974}, the renormalized energy associated with the spontaneous
density fluctuations that extend over a distance $\mathcal{\ell}_{qf}^{\ast}$
must be of the order $\left(\beta_{c}\right)^{-1}$, leading to a
renormalized free energy density of order $\left[\beta_{c}\left(\alpha_{c}\right)^{d}\right]^{-1}$.
Along the critical isochore, this energy will be associated to $\mathcal{A}_{qf}^{*}\left(\mathcal{T}^{*}\right)$
of Eq. (\ref{Aqfstar power law}). The product $\mathcal{A}_{qf}^{*}\left(\mathcal{T}^{*}\right)\times\left(\mathcal{\ell}_{qf}^{\ast}\right)^{d}$
being a universal quantity, the relative quantum correction to the
renormalized singular free energy reads\begin{equation}
\mathcal{A}_{qf}^{*}\left(\mathcal{T}^{*}\right)=\left(\Lambda_{qe}^{*}\right)^{d}\mathcal{A}^{*}\left(\mathcal{T}^{*}\right)\label{Aqfstar}\end{equation}
 due to the Eq. (\ref{lqfstar}) for $\mathcal{\ell}_{qf}^{\ast}$.
$\mathcal{A}^{*}\left(\mathcal{T}^{*}\right)$ must be the renormalized
singular free energy already defined for non-quantum fluids such as\begin{equation}
\mathcal{A}^{*}=\beta_{c}\left(\alpha_{c}\right)^{d}\times\frac{\Delta A}{V}\label{Astar}\end{equation}

\begin{figure*}
\includegraphics{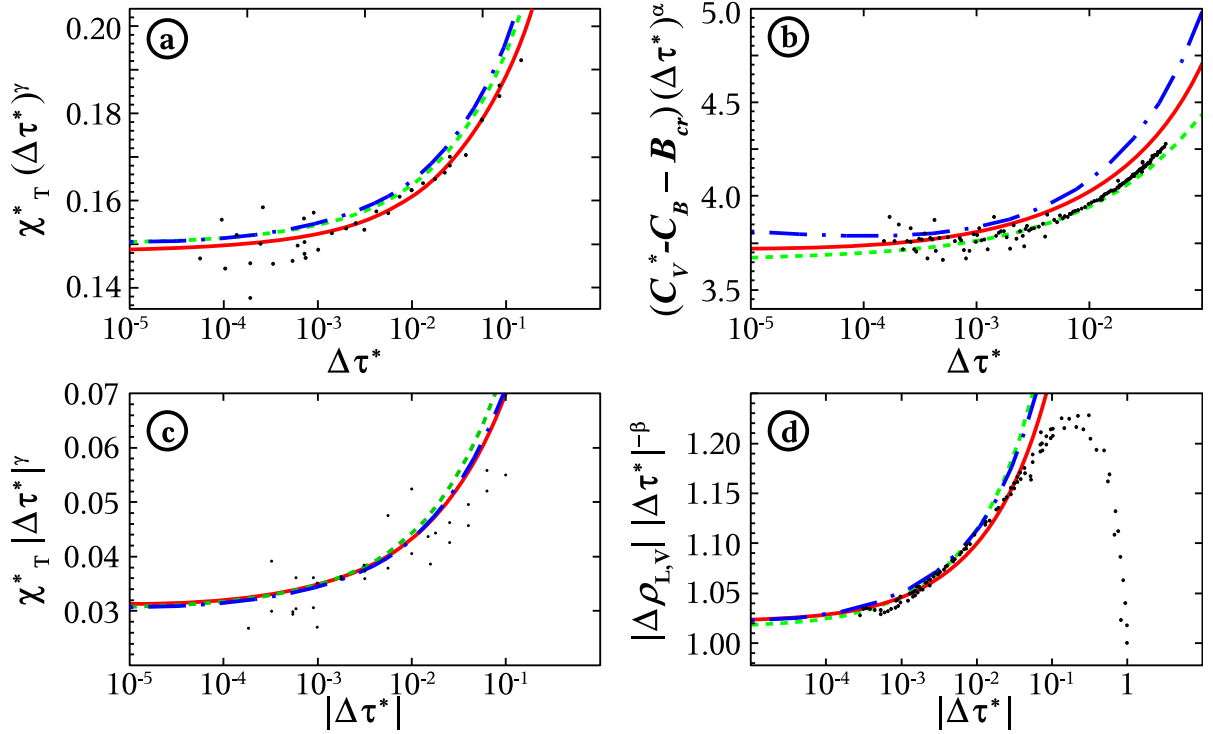}

\caption{Asymptotic two-term predictions compared to the $^{3}He$ measurements
(black points). The solid (red) lines are the actual predictions by
the dilated scale method (see Eq. (\ref{P fluid two term power law})
and column 4 of Table II). The dot-dashed (blue) lines are the two-term
prediction obtained from the best fit by the MSR $\Phi_{d=3}^{4}(1)$-model
\cite{Zhong2003} (Eq. (\ref{P fluid two term power law}); column
2 of Table II). The dashed (green) lines are the two-term prediction
obtained from CPM model \cite{Agayan2001} (Eq. (\ref{P fluid two term power law});
column 3 of Table II). (a) Susceptibility measurements for $T>T_{c}$
(corresponding to upper part of Fig. 1 in \cite{Zhong2003}). (b)
Specific heat measurements for $T>T_{c}$, where the small difference
in the additional constant term $C_{B}+B_{cr}$ is accounted in the
vertical scale (see also the lower part of Fig. 3 in \cite{Zhong2003}).(c)
Susceptibility measurements for $T<T_{c}$ (see also the lower part
of Fig. 1 in \cite{Zhong2003}). (d) Liquid-gas coexisting density
measurements (see also Fig. 4 in \cite{Zhong2003}).}
\end{figure*}

Therefore, from the comparison between the leading terms of the renormalized
and the physical second derivatives of the singular free energy densities
with respect to their associated thermal fields, we obtain 

\begin{equation}
A^{\pm}=\frac{1}{Z_{c}}\frac{A_{0,\bar{p}}^{\pm}}{k_{B}}=\left(\Lambda_{qe}^{*}\right)^{-d}\left(Y_{c}\right)^{2-\alpha}\mathcal{Z}_{\mathcal{C}}^{\pm}\label{A amplitude}\end{equation}
 In addition to the explicit $Y_{c}$ and $\Lambda_{qe}^{*}$ dependences
of the leading dimensionless amplitude $A^{\pm}$, the above Eqs.
(\ref{A amplitude}), also show the role of the particle number $\frac{1}{Z_{c}}$
such as the multiplicative factor to the leading particle amplitude
$A_{0,\bar{p}}^{\pm}\sim\left[k_{B}\right]$. That provides understanding
of the master (i.e. unique) singular behaviors of the one-component
fluid subclass in terms of the master (i.e. constant) properties of
the critical interaction cell of any one-component fluid. Similarly,
from the comparison between the leading terms of the renormalized
and the physical correlation lengths, we obtain\begin{equation}
\xi_{0}^{\pm}=\alpha_{c}\Lambda_{qe}^{*}\left(Y_{c}\right)^{-\nu}\mathcal{Z}_{\xi}^{\pm}\label{ksi amplitude}\end{equation}
In Eqs. (\ref{A amplitude}) and (\ref{ksi amplitude}), $A_{0,\bar{p}}^{\pm}\sim\left[k_{B}\right]$
and $\xi_{0}^{\pm}\sim\left[length\right]$ have the appropriate $Q_{c}^{min}$
and $\Lambda_{qe}^{*}$ dependences to satisfy the universal amplitude
combination of Eq. (\ref{universal combination Rksi}). These two
equations (\ref{A amplitude}) and (\ref{ksi amplitude}), or more
generally, all the relations given in the column 6 of Table I, also
demonstrate that the estimation of the adjustable parameter $\lambda_{q,f}$,
introduced throughout the Eqs. (\ref{lambdaqestar}) and (\ref{lambdaqf vs lambdac}),
is unequivocally made from the leading power law behavior of any property,
when $Q_{c}^{min}$ is known. That provides a very sensitive test
of the above phenomenological approach to account for quantum effects,
provided that the same length scale $\alpha_{c}$ and the same energy
scale $\left(\beta_{c}\right)^{-1}$ are used for thermodynamic and
correlation functions at $T\cong T_{c}$. In such a coherent thermodynamic
normalization, the relative quantum modification {[}proportional to
$\left(\Lambda_{qe}^{*}\right)^{d}$ {]} of the energy within the
critical interaction cell is correlated to the relative quantum modification
of the microscopic wave number {[}proportional to $\Lambda_{qe}^{*}${]}.
We thus provide the microscopic quantum mechanical modification which
complement the Widom's \cite{Widom1974} and Staufer et al's \cite{Staufer1972}
macroscopic argument, when it is expected that the free energy associated
to fluctuations of size $\xi$ were solely responsible for the singular
contribution of thermodynamic potentials and correlation functions.

\subsection{${}^{3}He$ results}

For the fermionic quantum fluid ${}^{3}He$, the $Q_{c}^{min}$ set
is composed of the following critical coordinates $T_{c}=3.315546\: K$,
$p_{c}=1.14724\,10^{5}\: Pa$, $\rho_{c}=41.45\: kg\, m^{-3}$, and
$\gamma_{c}^{'}=1.1759\,10^{5}\: Pa\, K^{-1}$ \cite{Zhong2003}.
Using Eqs. (\ref{length unit}) to (\ref{isochoric factor}), the
values of the four scale factors are $\left(\beta_{c}\right)^{-1}=4.5776\,10^{-23}J$,
$\alpha_{c}=7.362\,10^{-10}m$, $Y_{c}=2.39837$, $Z_{c}=0.301284$.
By $\chi^{2}$-optimization only using the susceptibility data above
and below $T_{c}$ in the range $\left|\Delta\tau^{*}\right|<5\,10^{-3}$,
with $\frac{\Gamma^{+}}{\Gamma^{+}}=\frac{\mathcal{Z}_{\mathcal{\chi}}^{+}}{\mathcal{Z}_{\mathcal{\chi}}^{-}}=4.79$
\cite{Guida1998}, the adjustable parameter {\small $\lambda_{q,f}$}
takes the numerical value {\small $\lambda_{q,{}^{3}He}=0.146423$},
leading to $\Lambda_{qe}^{*}=1.11009$. For the specific case of the
heat capacity, the additional critical $(B_{cr}^{*})$ and background
$(C_{B}^{*})$ terms are treated as one single adjustable constant
($B_{cr}^{*}+C_{B}^{*}$).

\begin{table*}
\begin{center}\begin{tabular}{|c|c|c|c|c|c|c|c|}
\hline 
Amplitude &
MSR \cite{Zhong2003}&
MSR \cite{Zhong2004}&
 MR6\cite{Zhong2004}&
 MR7\cite{Zhong2004}&
 CPM \cite{Agayan2001}&
 CPM \cite{Zhong2004}&
 This work \tabularnewline
\hline
$\xi_{0}^{+}\,\left(\overset{\circ}{A}\right)$&
 $2.71\pm0.02$&
&
&
&
 $2.68\pm0.04$&
&
 $2.68541$\tabularnewline
 $a_{\xi}^{+}$&
 $0.732\pm0.007$&
&
&
&
&
&
 $0.58474$\tabularnewline
 $\Gamma^{+}$&
 $0.150\pm0.007$&
 $0.147\pm0.001$&
 $0.146\pm0.001$&
 $0.148\pm0.001$&
 $0.150\pm0.002$&
 $0.153\pm0.001$&
 $0.148247$\tabularnewline
 $\Gamma^{-}$&
 $0.0303\pm0.0015$&
$0.0299\pm0.0003$&
 $0.0308\pm0.0001$&
 $0.0310\pm0.0001$&
&
 $0.0310\pm0.0002$&
 $0.030953$\tabularnewline
 $a_{\chi}^{+}$&
 $0.98\pm0.08$&
 $1.10\pm0.01$&
 $1.13\pm0.01$&
 $1.17\pm0.01$&
 $0.941\pm0.007$&
 $0.81\pm0.01$&
 $0.860931$\tabularnewline
 $a_{\chi}^{-}$&
 $4.29\pm0.34$&
 $4.83\pm0.05$&
 $3.58\pm0.05$&
 $5.30\pm0.07$&
&
 $4.17\pm0.07$&
 $4.01366$\tabularnewline
 $\frac{A^{+}}{\alpha}$&
 $3.73\pm0.45$&
 $3.76\pm0.05$&
 $3.72\pm0.01$&
 $3.84\pm0.02$&
 $3.548\pm0.031$&
 $3.63\pm0.02$&
 $3.71132$\tabularnewline
 $\frac{A^{-}}{\alpha}$&
$6.97\pm0.83$&
 $7.03\pm0.10$&
$6.883\pm0.026$&
$7.149\pm0.027$&
$6.823\pm0.01$&
$6.935\pm0.04$&
$6.90948$\tabularnewline
 $\alpha a_{C}^{+}$&
 $1.2\pm0.1$&
 $0.99\pm0.01$&
$1.13\pm0.01$&
$1.07\pm0.01$&
 $0.712\pm0.006$&
$0.61\pm0.01$&
 $0.810892$\tabularnewline
 $\alpha a_{C}^{-}$&
 $1.1\pm0.1$&
 $0.92\pm0.01$&
$1.17\pm0.01$&
$0.83\pm0.01$&
$0.593\pm0.012$&
$0.74\pm0.01$&
$0.59712$\tabularnewline
 $B_{cr}+C_{B}$&
 $-1.65\pm0.85$&
 $-1.67\pm0.13$&
 $-1.64\pm0.04$&
 $-1.81\pm0.04$&
 $-0.96\pm1.0$&
 $-1.23\pm0.05$&
 $-1.40$\tabularnewline
 $B$&
 $1.020\pm0.006$&
 $1.021\pm0.003$&
$1.008\pm0.004$&
$1.039\pm0.004$&
 $1.0047$&
$1.028\pm0.004$&
 $1.02134$\tabularnewline
 $a_{M}^{-}$&
 $0.91\pm0.02$&
 $0.91\pm0.01$&
$1.001\pm0.023$&
$0.218\pm0.003$&
 $0.8441$&
$0.73\pm0.01$&
 $0.77484$ \tabularnewline
\hline
\end{tabular}\end{center}

\caption{Calculated values for critical amplitudes of $^{3}He$ with $T_{c}=3.315546\: K$,
$p_{c}=1.14724\,10^{5}\: Pa$, $\rho_{c}=41.45\: kg\, m^{-3}$, $\gamma_{c}^{'}=1.1759\,10^{5}\: Pa\, K^{-1}$,
and $\lambda_{q,{}^{3}He}=0.146423$ (see text). Using Eqs. (\ref{length unit}),
(\ref{isochoric factor}) and the definition of $\Lambda_{qe}^{*}$
from Eqs. (\ref{lambdaqestar}) to (\ref{wavelength vs alphac}),
the values of the (five) characteristics parameters for $^{3}He$
are $\left(\beta_{c}\right)^{-1}=4.5776\,10^{-23}J$, $\alpha_{c}=7.362\,10^{-10}m$,
$Y_{c}=2.39837$, $Z_{c}=0.301284$, and $\Lambda_{qe}^{*}=1.11009$. }
\end{table*}

The main results are illustrated in Fig. 1 where the comparison is
made to the recent published experimental data (black points in Fig.
1) of Zhong et al \cite{Zhong2003}. In order to simplify the comparison,
the same corresponding scaled data by the asymptotic power law term
$\left(\Delta\tau^{\ast}\right)^{-x}$ were used for susceptibility
and heat capacity above $T_{c}$, and for coexisting liquid vapor
densities below $T_{c}$, which improves the sensitivity of the relative
representation from the asymptotic amplitude values. Obviously, that
provides a simultaneous significative test of the quantum effect contribution
since, among the four leading amplitudes $\Gamma^{+}$, $A^{+}$,
$\Gamma^{-}$, and $B$, only one is readily sufficient to define
the unequivocal $\Lambda_{qe}^{*}$-dependence. Moreover, to illustrate
the first confluent term contribution associated to the scale dilatation
method, the full (red) lines and the dot-dashed (blue) lines in Fig.
1 correspond to the respective first-order Wegner expansions obtained
from Table I, and from the Zhong et al \cite{Zhong2003} initial fit,
using the minimal substraction renormalization (labelled $MSR$) scheme
(see also the corresponding numerical values of the amplitudes listed
in Table II). For the three selected properties, the predicted singular
behavior fits well the experimental results and matches the theoretical
predictions of the minimal-substraction renormalization scheme.

More generally, as shown in Table II, the two-term asymptotical results
obtained with the scale dilatation method are in good agreement with
the two-term parametric modeling recently obtained by Zhong and Barmatz
\cite{Zhong2004}, based on three different theoretical models. Two
of these models are issued from the two main field-thoretical renormalization
schemes that treat classical-to-critical crossover phenomena, namely
the minimal-substraction renormalization scheme of Dombs and co-workers
\cite{Dohm1985,Schloms1987,Schloms1989,Schloms1990,Larin1998}, and
the massive renormalization scheme of Bagnuls and Bervillier \cite{Bagnuls1984a,Bagnuls1985,Bagnuls1987,Bagnuls2002},
only applied to the primary critical path corresponding to the homogeneous
and non homogeneous domain along the critical isochore. The third
model, namely the crossover parametric model, proposed by Agayan and
coworkers \cite{Agayan2001,Barmatz2000}, is a complete parametric
equation of state issued from a phenomenological crossover transformation
of a classical Landau expansion of the singular free energy \cite{Chen1990a,Chen1990b,Anisimov1992}.
Although it is phenomenological, this crossover Landau model was successfully
applied to several one-component fluids. A previous comparison of
the results obtained by the crossover Landau model and the scale dilatation
method was already made in the case of seven non-quantum fluids. In
Table II are reported,

i) colums 2 and 3 labelled $MSR$, the results obtained by Zhong et
al \cite{Zhong2003} and Zhong and Barmatz \cite{Zhong2004} from
the minimal-substraction renormalization scheme;

ii) colums 4 and 5 labelled $MR6$ and $MR7$, the results obtained
by Zhong and Barmatz \cite{Zhong2004} from the massive renormalization
scheme in the sixth- \cite{Bagnuls1985,Bagnuls1987} and seventh-loop
\cite{Bagnuls2002} series;

iii) colums 6 and 7 labelled $CPM$, the results obtained by Agayan
et al \cite{Agayan2001} and Zhong and Barmatz \cite{Zhong2004} from
the crossover parametric model;

iv) column 8, the results obtained in this work applying the scale
dilatation method.

The excellent agreement between the amplitude values permits to discuss
now the introduction of the adjustable parameters in the modeling,
and to explain why only two ajustable parameters in the models are
significant with respect to the fit quality, as concluded by Zhong
and Barmatz \cite{Zhong2004}.

\section{${}^{3}He$ critical modelling}

\subsection{The two renormalization schemes along the critical isochore}

As clearly mentionned in the Appendix D of the reference \cite{Zhong2003},
the three free parameters of the $MSR$-model originate from the undetermined
integration constants $z_{\phi}$, $z_{a}$, and $z_{\mu}$, associated
to the flow equations of their respective $Z_{\phi}\left(u\right)$,
$Z_{r}\left(u\right)$, and $Z_{u}\left(u\right)\left[Z_{\phi}\left(u\right)\right]^{-2}$
field theoretical functions (here we have adopted the Zhong et al
notation for $z_{a}$ and $z_{\mu}$, adding $z_{\phi}$ as being
the undetermined integration constant to solve Eq. (7) of reference
\cite{Zhong2003}). These integration constants are system dependent
and can be obtained by fitting experimental data to the theory. However,
considering uniquely the critical isochore, the given set composed
by the explicit adjustable parameters (such as $\left\{ u,\mu,a\right\} $
in the MSR model case \cite{Hahn2001,Zhong2003}) or calculated parameters
(such as the leading amplitudes and $t_{0}$ in the MSR model case
\cite{Hahn2001,Zhong2003}) result in complicated scaled forms of
combinations between $z_{\phi}$, $z_{a}$, and $z_{\mu}$. Specifically,
to account correctly for the $z_{\phi}$ system dependence needs to
use several properties. The susceptibility fitting results reported
in Figure 2 of Ref. \cite{Zhong2003}, where only two ($\mu$ and
$t_{0}$) among the three scaled parameters ($\mu$, $a$ and $t_{0}$)
have the expected power law dependence on $1-\frac{u}{u^{*}}$ near
the fixed point ($u=u^{*}$), should be also due to a non-representative
test of one asymptotical scaled form. A preliminary comparison of
the functional forms of the leading amplitudes obtained from the MSR
model and the scale dilatation method for the case of the non-quantum
fluid subclass, suggests for example that the true independent scaled
factors of each physical system are then such as\begin{equation}
\frac{z_{a}}{\left(z_{\mu}\right)^{\zeta_{r}^{*}}}\propto Y_{c}\label{zaoverzmu vs Yc}\end{equation}
 and\begin{equation}
\frac{z_{\phi}}{\left(z_{\mu}\right)^{\zeta_{\phi}^{*}}}\propto Z_{c}\label{zphioverzmu vs Zc}\end{equation}
 In Eqs. (\ref{zaoverzmu vs Yc}) and (\ref{zphioverzmu vs Zc}),
$\zeta_{r}^{*}=\zeta_{r}\left(u^{*}\right)=1-\frac{1}{\nu}$ and $\zeta_{\phi}^{*}=\zeta_{\phi}\left(u^{*}\right)=-\eta$
are the respective values of the field theoretical functions at the
Ising fixed point $u=u^{*}$ (see Zhong et al's \cite{Zhong2003}
notations). This suggestion should be usefull for a possible rescaling
of the leading amplitudes which gives better evidence for the two
asymptotical parameters which are readily independent in the modeling
form the minimal-substraction renormalization scheme.

The two-term master asymptotical behavior obtained from the scale
dilatation method can be described \cite{Garrabos2005a} by the massive
renormalization scheme of Bagnuls and Bervillier, thanks to its formal
analogy to the basic analytical hypotheses of the renormalization
\cite{Wilson1971,Wilson1974}. Using a similar approach which introduces
one common (i.e. $P^{*}$-independent) crossover parameter $\vartheta_{{}^{3}He}$,
and adjustable prefactors $\mathbb{P}_{0,{}^{3}He}^{\pm}$ for each
dimensionless property $P^{*}$, we obtain the following values $\mathbb{L}_{0,{}^{3}He}^{+}=1.2925$
, $\mathbb{X}_{0,{}^{3}He}^{+}=1.818$, $\mathbb{C}_{0,{}^{3}He}^{+}=2.1503$,
and $\mathbb{M}_{0,{}^{3}He}^{\pm}=1.0894$, for the leading prefactors
of the correlation length, the susceptibility, the heat capacity and
the coexistence curve, respectively. These four leading parameters
are interrelated by the following combinations $\mathbb{L}_{0,{}^{3}He}^{+}\left(\mathbb{C}_{0,{}^{3}He}^{+}\right)^{\frac{1}{d}}=1$
and $\frac{\mathbb{X}_{0,{}^{3}He}^{+}}{\left(\mathbb{M}_{0,{}^{3}He}^{\pm}\right)^{2}}\left(\mathbb{L}_{0,{}^{3}He}^{+}\right)^{-d}=1$,
so that only two of them are independent, by virtue of the two scale
factor universality. The estimated value of the crossover parameter
is $\vartheta_{{}^{3}He}=0.0113$. The mean crossover functions \cite{Garrabos2005b}
will be used in a future work to implement the master estimation of
their free parameters from the four scale factors defined by $Q_{c}^{min}$.

\subsection{The crossover parametric model of the e.o.s.}

The crossover parametric model is issued from the crossover Landau
model (CLM) of the e.o.s. based on a phenomenological transformation
of a classical Landau expansion of the singular free energy of the
fluid as a function of the local order parameter density. In such
a modeling, the simplest crossover description involves three free
parameters, made of the two coupling constants $a_{0}$ and $u_{0}$
and one gradient prefactor $c_{0}$ (see for example \cite{Agayan2001}
for notations). After transformation of variables and coefficients,
the three initial system-dependent coefficients $a_{0}$, $u_{0}$,
and $c_{0}$, are replaced by two dimensionless asymptotic scaling
parameters (noted $c_{t}$ and $c_{\rho}$ in the general CLM approach)
and one dimensionless crossover parameter (noted $g$). However, from
the field theory framework, any description of a 3D Ising like system
with {}``finite'' cutoff, needs to maintain the appropriate interdependence
between the nonuniversal parameters, specially the microscopic wavelength
$\Lambda_{0}\sim\left[length\right]^{-1}$ and the coupling constant
$u_{0}\sim\left[length\right]$. Introducing then a common \emph{arbitrary}
length scale unit permits to replace the product $u_{0}\Lambda_{0}$
by the product $u\Lambda$ of the corresponding dimensionless wavelength
$\Lambda$ and coupling constant $u$. The convenient normalization
$\bar{u}=\frac{u}{u^{*}}$, where $u^{*}$ corresponds to the universal
value at the non Gaussian fixed point, leads to an arbitrary choice
for the dimensionless microscopic wavelength $\Lambda$ and the dimensionless
coupling parameter $\bar{u}$, provided that $\bar{u}\Lambda$ remains
finite in order to account for theoretical infinite cutoff approximation,
$\Lambda\rightarrow\infty$ and $\bar{u}\rightarrow0$. In this infinite-cutoff
limit where $g$ is related to the Ginzburg number \cite{Anisimov1992},
the crossover behavior is then universal by rescaling the thermal
field like variable using a single crossover parameter (such as $g=\frac{\left(\bar{u}\Lambda\right)^{2}}{c_{t}}=\Delta\tau_{X}^{*}$,
or such as the crossover temperature $t_{X}=c_{t}\Delta\tau_{X}^{*}$,
equivalently \cite{Anisimov1992}), However, at the general symmetrical
fourth-order (with only two independent coupling quantities $a_{0}$
and $u_{0}$) of the phenomenological transformation of the classical
Landau expansion of the singular free energy, the crossover behavior
is governed by the two dimensionless parameters $g$ and $\bar{u}$.
In such a situation all the dimensionless quantities are canonical
constants, provided one have defined a microscopic characteristic
length scale for each fluid. That provides implicit connection between
$\Lambda$ and $\bar{u}$, or equivalently between $\Lambda$ and,
for example $c_{t}$, when the explicit $g=\frac{\left(\bar{u}\Lambda\right)^{2}}{c_{t}}$
dependence is accounted for, as mentionned above. As a consequence,
the only way to monitor the asymptotic critical behavior of the crossover
Landau model is to change $\bar{u}$, or equivalently $c_{t}$. We
recall that, in a previous analysis of the corresponding results for
the case of seven non-quantum fluids \cite{Garrabos2002}, we have
shown that\begin{equation}
c_{t}\left(\bar{u}\Lambda\right)=Y_{c}\times f_{t}\left(\bar{u}\Lambda\right)\label{ct vs Yc}\end{equation}
 and\begin{equation}
c_{\rho}\left(\bar{u}\Lambda\right)=\left(Z_{c}\right)^{\frac{1}{2}}\times f_{\rho}\left(\bar{u}\Lambda\right)\label{crho vs Zc}\end{equation}
 are unequivocally well-related to our scale factors $Y_{c}$ and
$Z_{c}$, respectively. In Eqs. (\ref{ct vs Yc}) and (\ref{crho vs Zc}),
$f_{t}\left(\bar{u}\Lambda\right)$ and $f_{\rho}\left(\bar{u}\Lambda\right)$are
two appropriate universal power laws of the product $\bar{u}\Lambda$,
uniquely. In the following, we will also provide one possible estimation
of the coupling constants $a_{0}\left(g\right)$ and $u_{0}\left(g\right)$
from $Y_{c}$ and $Z_{c}$, now using the system-dependent coefficients
of the crossover parametric model.

The \emph{three-parameter} crossover parametric model contains two
asymptotic scaling parameters, noted $l_{0}$ and $m_{0}$, and again
the crossover parameter $g$. A comparison between definitions of
asymptotic amplitudes $\Gamma^{+}$ and $B$ leads to the following
relations,\begin{equation}
l_{0}=\frac{3.38317}{\mathcal{Z}_{\mathcal{X}}}\frac{\mathcal{Z}_{\mathcal{M}}}{3.28613}\left(\Lambda_{qe}^{*}\right)^{-2}\left(Z_{c}\right)^{\frac{1}{2}}\left(Y_{c}\right)^{\beta+\gamma}\label{lzero}\end{equation}
 and\begin{equation}
m_{0}=\frac{\mathcal{Z}_{\mathcal{M}}}{3.28613}\left(\Lambda_{qe}^{*}\right)^{-1}\left(Z_{c}\right)^{-\frac{1}{2}}\left(Y_{c}\right)^{\beta}\label{mzero}\end{equation}
(see our Table I and Table III of Ref. \cite{Agayan2001} for details).
Our direct estimation of the two free values $l_{0}=7.0929$ and $m_{0}=0.3108$
from Eqs. (\ref{lzero}) and (\ref{mzero}), are in close-agreement
with the values $l_{0}=6.89\pm0.12$ and $m_{0}=0.306\pm0.01$, deduced
from the fitting procedure of Agayan et al (see Ref. \cite{Agayan2001}),
and with the values $l_{0}=6.902\pm0.012$ and $m_{0}=0.3128\pm0.0004$,
recently obtained by Zhong and Barmatz \cite{Zhong2004} in their
recent comparison of theoretical models of crossover behavior. Moreover,
as previously mentionned, from the identification of the leading amplitudes
given in Table I of Ref. \cite{Agayan2001}, calculated using, either
the crossover Landau model, or the crossover parametric model, it
is now easy to show that the two coupling constants $a_{0}$ and $u_{0}$
are related to $Y_{c}$ and $Z_{c}$ (and $\Lambda_{qe}^{*}$, obviously),
by the following relations\begin{equation}
a_{0}\left(g\right)=\left(\Lambda_{qe}^{*}\right)^{-1}Z_{c}\left(Y_{c}\right)^{\gamma}f_{a_{0}}\left(g\right)\label{azero vs Qcmin scale factors}\end{equation}
 and\begin{equation}
u_{0}\left(g\right)=\Lambda_{qe}^{*}\left(Z_{c}\right)^{2}\left(Y_{c}\right)^{2\beta-\gamma}f_{u_{0}}\left(g\right)\label{uzero vs Qcmin scale factors}\end{equation}
 In Eqs. (\ref{azero vs Qcmin scale factors}) and (\ref{uzero vs Qcmin scale factors}),
$f_{a_{0}}\left(g\right)$ and $f_{u_{0}}\left(g\right)$ are two
appropriate universal power laws of the crossover parameter $g$. 

The first confluent amplitude for the susceptibility obtained from
the crossover parametric model reads $\Gamma_{1}^{+}=g_{\chi}^{+}g^{-\Delta_{s}}\left(1-\bar{u}\right)$,
with $g_{\chi}^{+}=0.590$, $\Delta_{s}=0.51$, and $g=\frac{\left(\bar{u}\Lambda\right)^{2}}{c_{t}}$
(see Table III of Ref. \cite{Agayan2001}). The identification with
our corresponding amplitude $a_{\chi}^{+}=\mathcal{Z}_{\chi}^{1,+}\left(Y_{c}\right)^{\Delta}$
(Table 1), gives\begin{equation}
g_{\chi}^{+}\left(\frac{\bar{u}\Lambda}{\left(c_{t}\right)^{\frac{1}{2}}}\right)^{-2\Delta_{s}}\left(1-\bar{u}\right)=\mathcal{Z}_{\chi}^{1,+}\left(Y_{c}\right)^{\Delta}\label{gubarlambda vs Yc}\end{equation}
 demonstrating unequivocal relation between $g^{\frac{1}{2}}=\frac{\bar{u}\Lambda}{\left(c_{t}\right)^{\frac{1}{2}}}$
and $Y_{c}$. However, the rescaled coupling constant $\bar{u}$ remains
dependent, on the one hand, to the correlation between the three adjustable
dimensionless parameters $c_{t}$, $\bar{u}$, and $\Lambda$ of the
model, and on another hand, to the master value $\mathcal{Z}_{\chi}^{1,+}=0.555$
initially estimated from the analysis of the isothermal compressibility
data of xenon. That implies the implicit introduction of one characteristic
microscopic length which must take a unique {}``thermodynamic''
definition {[}by Eq. (\ref{length unit}){]}, whatever the selected
one-component fluid. In that {}``normalized'' situation, our present
value $a_{\chi}^{+}=0.861$ for $^{3}He$, results in good agreement
with for example the values $\Gamma_{1}^{+}=0.941\pm0.007$ \cite{Agayan2001}
and $\Gamma_{1}^{+}=0.81\pm0.01$ \cite{Zhong2004} obtained from
data fitting with the crossover parametric model (see below for more
details on the uncertainty associated to the $\Gamma_{1}^{+}$ determination).
Accounting for the arbitrary relation $\frac{\Lambda}{\left(c_{t}\right)^{\frac{1}{2}}}=\pi$
adopted by the authors of Ref. \cite{Agayan2001}, our calculated
value $\bar{u}=0.18075$ from Eq. (\ref{gubarlambda vs Yc}), yields
to $g^{\frac{1}{2}}=0.5678$, which compares favourably to $g^{\frac{1}{2}}=0.528\pm0.003$
obtained from the data fitting performed by Agayan et al \cite{Agayan2001}.
Accounting for the arbitrary relation $\frac{\Lambda}{\left(c_{t}\right)^{\frac{1}{2}}}=\frac{\pi}{\sqrt{6}}$
adopted by the authors of Ref. \cite{Zhong2004}, with $\Lambda$
fixed at unity (yielding to $g^{\frac{1}{2}}=u^{*}=0.472)$, our calculated
value $\bar{u}=0.35187$ from (\ref{gubarlambda vs Yc}) (yielding
to $g^{\frac{1}{2}}=0.4513$), compares favourably to $\bar{u}=0.368\pm0.004$
obtained from the Zhong et al \cite{Zhong2004} data fitting. The
$\sim10$ \% residual difference between these two estimations of
the fluid-dependent parameters, reflects the small differences between
theoretical values of universal exponents and amplitude combinations,
added to the uncertainty in the direct estimation of the confluent
amplitude, the latter one being greater than $10$\% (for example,
using {}``equivalent'' crossover Landau modeling of the same $^{3}He$
experimental data, the resulting values are $\Gamma_{1}^{+}=0.946\pm0.006$
and $\Gamma_{1}^{+}=1.000\pm0.028$ in Ref. \cite{Barmatz2000}, $\Gamma_{1}^{+}=0.941\pm0.007$
in Ref. \cite{Agayan2001}, and $\Gamma_{1}^{+}=0.81\pm0.01$ in Ref.
\cite{Zhong2004}, while using the minimal-substraction renormalization
scheme, the resulting values are $\Gamma_{1}^{+}=1.01\pm0.08$ in
Ref. \cite{Hahn2001}, $\Gamma_{1}^{+}=0.98\pm0.08$ or $\Gamma_{1}^{+}=1.13\pm0.01$
in Ref. \cite{Zhong2003}, and $\Gamma_{1}^{+}=1.10\pm0.01$ in Ref.
\cite{Zhong2004}, leading to the practical {}``mean'' value $\Gamma_{1}^{+}=0.97\pm0.16$)
(see also Table II). Nevertheless, this agreement confirms our previous
analyses \cite{Bagnuls1984b,Garrabos2002} of the confluent correction
to scaling for the one-component fluid subclass satisfying to the
classical-to-critical crossover description along the ideal RG trajectory
\cite{Bagnuls1997,Bagnuls2000}.

\section{Conclusions}

The present study in terms of the dilated physical fields for quantum
fluids adds only one well-defined adjustable parameter, which accounts
for microscopic quantum effects only asymptotically close to the critical
point ($T\cong T_{c}$). The adjustable parameter is introduced in
a phenomenological manner which maintains universal feature of the
singular free energy in a appropriate microscopique volume. Since
our selected standard fluid is xenon, we provide here a complementary
new light to the recent discussions \cite{Luijten2000,Hahn2001} about
the definitions of the crossover temperature $t_{X}$ {[}related to
the crossover parameter $g$, (or the Ginzburg number), as mentionned
in §.4.2{]}. As an essential new consequence, we note that $t_{X}\propto\frac{1}{Y_{c}}$
along the critical isochore, for $T>T_{c}$. Therefore, our two-term
asymptotic hyperscaling seems also compatible with (at least) the
first-order contribution to the critical crossover. However this observed
supplementary constraint is not a necessity from the field theory
framework \cite{Bagnuls2002}. Consequently, our next work \cite{Garrabos2005c}
is to provide thermodynamic fundaments for the asymptotic master behavior
of thermodynamic and correlation functions which was inferred from
the above minimal information. In addition to the derivation of such
thermodynamic fundaments, we also propose a convenient mean form \cite{Garrabos2005a,Garrabos2005b}
of the max and min forms for each complete crossover function recently
derived by Bagnuls and Bervillier \cite{Bagnuls2002}. Such means
functions can be appropriately modified to account for the results
obtained by the scale dilatation method \cite{Garrabos2005a}, extending
thus the analysis of the crossover behavior of the one-component fluids
outside their Ising-like preasymptotic domain.

\section{Acknowledgments}

This work has benifited from a close collaboration with C. J. Erkey,
C. Lecoutre, B. Leneindre, F. Palencia and many stimulating discussions
with C. Bagnuls and C. Bervillier. We are also grateful to M. Barmatz
and F. Zhong for enlightening comments concerning their experimental
results and theoretical analyses.


\begin{thebibliography}{10}
\bibitem{Anisimov2000}M. A. Anisimov and J. V. Sengers, in {}``\emph{Equations of State
for Fluids and Fluid Mixtures}'', Part I, J. V. Sengers, R. F. Kayser,
C. J. Peters, and H. J. White, Jr., Eds. (Elsevier, Amsterdam, UK,
2000), pp. 381-434. 
\bibitem{ZinnJustin2002}See for example J. Zinn Justin, \emph{Quantum Field Theory and Critical
Phenomena}, $4^{th}$ ed. (Oxford University Press, 2002).
\bibitem{Wegner1972}F. J. Wegner, Phys. Rev. B \textbf{5,} 4529 (1972).
\bibitem{Bagnuls1984a}C. Bagnuls and C. Bervillier, J. Phys.-\textit{Lettres} \textbf{45},
L-95 (1984). 
\bibitem{Bagnuls1985}C. Bagnuls and C. Bervillier, Phys. Rev. B \textbf{32,} 7209 (1985). 
\bibitem{Kim2003}Y. C. Kim, M. E. Fisher, and G. Orkoulas, Phys. Rev. E \textbf{67},
061506 (2003); and references cited therein.
\bibitem{Garrabos1982}Y. Garrabos, Thesis, University of Paris VI (1982), unpublished.
\bibitem{Garrabos1985}Y. Garrabos, J. Phys. \textbf{46}, 281 (1985) {[}for an english version
see https://hal.ccsd.cnrs.fr/ccsd-00015956 (15 Dec. 2005), or http://fr.arxiv.org/abs/cond-mat/0512347{]};
J. Phys. \textbf{47,} 197 (1986). 
\bibitem{Bagnuls1987}C. Bagnuls, C. Bervillier, D. I. Meiron, and B. G. Nickel, Phys. Rev.
B \textbf{35}, 3585 (1987). 
\bibitem{Bagnuls1984b}C. Bagnuls, C. Bervillier, and Y. Garrabos, J. Phys.-\textit{Lettres}
\textbf{45}, L-127 (1984).
\bibitem{aparticle notation}The subscript $a_{\bar{p}}$ in $Q_{c,a_{\bar{p}}}^{min}$ notation,
recalls for the Helmholtz free energy per particle $a_{\bar{p}}\left(T,v_{\bar{p}}\right)=\frac{A\left(T,\frac{V}{N},1\right)}{N}$
where $A\left(T,V,N\right)$ is the total Helmholtz free energy of
the fluid and $T$ (temperature), $V$ (total volume), $N$ (total
temperature), its associated three natural variables. $v_{\bar{p}}=\frac{V}{N}$
is the particle volume. From the general point of vue of the thermodynamics,
the equilibrium states of a one-component system at constant amount
of matter $N$, are entirely defined by the knowledge of the normalized
potential $a_{\bar{p}}\left(T,v_{\bar{p}}\right)$ or, alternatively
but equivalenty, by the knowledge of the two equations of state $s_{\bar{p}}\left(T,v_{\bar{p}}\right)=\left(\frac{\partial a_{\bar{p}}}{\partial T}\right)_{v_{\bar{p}}}$
and $p\left(T,v_{\bar{p}}\right)=\left(\frac{\partial a_{\bar{p}}}{\partial v_{\bar{p}}}\right)_{T}$.
$s_{\bar{p}}$ is the entropy of the particle and $p$ is the pressure.
These equilibrium states are represented by one 3D characteristic
surface of equation $\Phi(a_{\bar{p}},T,v_{\bar{p}})=0$, or alternatively
but equivalently, by two 3D phase surfaces of equations $\Phi(s_{\bar{p}},T,v_{\bar{p}})=0$
and $\Phi(p,v_{\bar{p}},T)=0$, respectively. Only the latter phase
surface can be constructed from $p,V,T,M$(total mass) measurements,
when the mass $m_{\bar{p}}$ of the particle is known {[}with $M=Nm_{\bar{p}}${]}.
\bibitem{Hirschfelder1954}See for example J. O. Hirschfelder, C. F. Curtiss, and R. B. Bird,
\emph{Molecular Theory of Gases and Liquids} (Wiley, New York, 1954).
\bibitem{Hahn2001}I. Hahn, F. Zhong, M. Barmatz, R. Haussmann, and J. Rudnick, Phys.
Rev. E \textbf{52}, 055104(R) (2001).
\bibitem{Zhong2003}F. Zhong, M. Barmatz, and I. Hahn, Phys. Rev. E \textbf{67}, 021106
(2003).
\bibitem{preferred}This topologic consideration agrees with the ideas of Griffiths and
Wheeler {[}R. B. Griffiths and J. C. Wheeler, Phys. Rev. A \textbf{2},
1047 (1970){]} concerning the preferred directions in the space of
the independent field variables.
\bibitem{Dchapeau}Using standard notations for correlation amplitudes $\widehat{D}$
and $\xi_{0}^{+}$; see for example \cite{Privman 1991} and also
\cite{Garrabos1985} for the one-component fluid case.
\bibitem{Campbell1978}See for example C. E. Campbell, in {}``Progress in Liquid Physics'',
Ed. C. A. Croxton (Wiley, New York, 1978) Chapter 6, pp. 213-308.
\bibitem{Thirumalai1983}D. Thirumalai, E. J. Bruskin, and B. J. Berne, J. Chem. Phys. \textbf{79,}
5063 (1983). 
\bibitem{Pollock1984}E. L. Pollock and D. M. Ceperley, Phys. Rev. B, \textbf{30} 2555 (1984). 
\bibitem{Muser2002}M. H. M$\ddot{u}$ser and E. Luijten, J. Chem. Phys. \textbf{116},
1621 (2002).
\bibitem{Garrabos2002}Y. Garrabos, B. Le Neindre, R. Wunenburger, C. Lecoutre-Chabot, and
D. Beysens, Int. J. Thermophys. \textbf{23,} 997 (2002).
\bibitem{Wilson1971}K. G. Wilson, Phys. Rev. B 4, 3174 (1971).
\bibitem{Wilson1974}K. G. Wilson and J. Kogut, Phys. Rep. C 12, 75 (1974).
\bibitem{Bagnuls2002}C. Bagnuls and C. Bervillier, Phys. Rev. E \textbf{65}, 066132 (2002).
\bibitem{Guida1998}R. Guida and J. Zinn-Justin, J. Phys. A: Math. Gen. \textbf{31}, 8103
(1998).
\bibitem{Sengers1986}J. V. Sengers and J. M. H. Levelt Sengers, Annu. Rev. Phys. Chem.
\textbf{37}, 189, (1986).
\bibitem{Privman 1991}V. Privman, P. C. Hohenberg, and A. Aharony, in {}``\emph{Phase Transitions
and Critical Phenomena}'', Ed. C. Domb and M. S. Green, Vol. 14 (Academic
Press, New York, 1991) Chapter 1, pp. 1-134.
\bibitem{Widom1974}B. Widom, J. Chem. Phys. \textbf{43}, 3898 (1965); Physica \textbf{73,}
107 (1974). 
\bibitem{Staufer1972}D. Staufer, M. Ferer, and M. Wortis, Phys. Rev. Lett \textbf{29},
345 (1972).
\bibitem{Zhong2004}F. Zhong and M. Barmatz, Phys. Rev. E 7\textbf{0}, 066105 (2004).
\bibitem{Dohm1985}V. Dohm, Z. Phys. B: Condens. Matter \textbf{60}, 61 (1985).
\bibitem{Schloms1987}R. Schloms and V. Dohm, Europhys. Lett. \textbf{3}, 413 (1987).
\bibitem{Schloms1989}R. Schloms and V. Dohm, Nucl. Phys. B \textbf{328}, 639 (1989).
\bibitem{Schloms1990}R. Schloms and V. Dohm, Phys. Rev. B \textbf{42}, 6142 (1990).
\bibitem{Larin1998}S. A. Larin, M. Mönnigmann, M. Strösser, and V. Dohm, Phys. Rev. B
\textbf{58}, 3394 (1998).
\bibitem{Agayan2001}V. A. Agayan, M. A. Anisimov, and J. V. Sengers, Phys. Rev. E \textbf{64},
026125 (2001).
\bibitem{Barmatz2000}M. Barmatz, I. Hahn, F. Zhong, M. A. Anisimov, and V. A. Agayan, J.
Low Temp. Phys. \textbf{121}, 633 (2000).
\bibitem{Chen1990a}Z.Y. Chen, P. C. Albright, and J. V. Sengers, Phys. Rev. A \textbf{41},
3161 (1990).
\bibitem{Chen1990b}Z. Y. Chen, A. Abbaci, S. Tang, and J. V. Sengers, Phys. Rev. A \textbf{42},
4470 (1990).
\bibitem{Anisimov1992}M. A. Anisimov, S. B. Kiselev, J. V. Sengers, and S. Tang, Physica
A \textbf{188}, 487 (1992). 
\bibitem{Garrabos2005a}Y. Garrabos, F. Palencia, C. Lecoutre, C. J. Erkey, and B. Le Neindre,
preprint (2005), submitted to Phys. Rev. E.
\bibitem{Garrabos2005b}Y. Garrabos, F. Palencia, C. Lecoutre, and C. Bervillier, preprint
(2005).
\bibitem{Bagnuls1997}C. Bagnuls and C. Bervillier, J. Phys. Stud. \textbf{1}, 366 (1997).
\bibitem{Bagnuls2000}C. Bagnuls and C. Bervillier, Condens Matter Phys. \textbf{3}, 559
(2000).
\bibitem{Luijten2000}E. Luijten and H. Meyer, Phys. Rev. E \textbf{62}, 3257 (2000).
\bibitem{Garrabos2005c}Y. Garrabos, preprint (2005), submitted to Europhys. Letters. \end{thebibliography}
\end{document}